\documentclass[12pt,preprint]{aastex}
\usepackage{epsfig}
\usepackage{natbib}
\usepackage{graphicx}
\usepackage{mathrsfs,amssymb}
\usepackage{amsmath}
\newcommand	\beq	{\begin{equation}}	
\newcommand	\eeq	{\end{equation}}	
\newcommand       \Angstrom     {\,{\rm \AA}}

\newcommand       \cm           {\,{\rm cm}}

\newcommand       \g            {\,{\rm g}}

\newcommand       \Gyr      {\,{\rm Gyr}}

\newcommand       \nH           {n_{\rm H}}
\newcommand       \NH           {N_{\rm H}}
\newcommand       \simlt        {\lesssim}
\newcommand       \simgt        {\gtrsim}

\newcommand       \um           {\mu{\rm m}}
\newcommand       \mum          {\,{\rm \mu m}}
\newcommand       \ppm          {\,{\rm ppm}}

\newcommand       \mH           {m_{\rm H}}

\newcommand       \simali       {\sim\,}
\newcommand       \magni        {\,{\rm mag}}
\newcommand       \rmH          {{\rm H}}
\newcommand       \HH           {\,{\rm H}}
\newcommand       \rhosil        {\rho_{\rm sil}}
\newcommand       \rhogra        {\rho_{\rm gra}}
\newcommand       \rhoC          {\rho_{\rm C}}
\newcommand	  \xsun         {\left[{\rm X/H}\right]_{\odot}}
\newcommand	  \xism         {\left[{\rm X/H}\right]_{\rm ISM}}
\newcommand	  \csun         {\left[{\rm C/H}\right]_{\odot}}

\newcommand	  \fesun        {\left[{\rm Fe/H}\right]_{\odot}}
\newcommand	  \mgsun        {\left[{\rm Mg/H}\right]_{\odot}}
\newcommand	  \sisun        {\left[{\rm Si/H}\right]_{\odot}}
\newcommand	  \xstar         {\left[{\rm X/H}\right]_{\star}}
\newcommand	  \cstar         {\left[{\rm C/H}\right]_{\star}}

\newcommand	  \sistar        {\left[{\rm Si/H}\right]_{\star}}
\newcommand	  \cism         {\left[{\rm C/H}\right]_{\rm ISM}}

\newcommand	  \siism        {\left[{\rm Si/H}\right]_{\rm ISM}}

\newcommand	  \cdust        {\left[{\rm C/H}\right]_{\rm dust}}

\newcommand	  \sidust       {\left[{\rm Si/H}\right]_{\rm dust}}

\newcommand	  \cgas        {\left[{\rm C/H}\right]_{\rm gas}}

\newcommand	  \sigas       {\left[{\rm Si/H}\right]_{\rm gas}}
\newcommand	  \mux         {\mu_{\rm X}}
\newcommand	  \muc         {\mu_{\rm C}}
\newcommand	  \muo         {\mu_{\rm O}}

\newcommand	  \muh         {m_{\rm H}}
\newcommand	  \musil       {\mu_{\rm sil}}

\newcommand	  \Vsil         {V_{\rm sil}}

\newcommand	  \VC           {V_{\rm C}}
\newcommand	  \Fsil         {F_{\rm sil}}

\newcommand	  \FC           {F_{\rm C}}
\newcommand	  \Aint         {A_{\rm int}}

\newcommand       \Bsil          {B_{\rm sil}}
\newcommand       \Bgra          {B_{\rm gra}}

\def    \RV		{R_V}
\def    \AV		{A_V}
\def    \nh		{n_{\rm H}}
\def    \acC		{a_{c,{\rm C}}}
\def    \acS		{a_{c,{\rm S}}}
\def    \Cext		{C_{\rm ext}}
\def    \alphaC		{\alpha_{\rm C}}
\def    \alphaS		{\alpha_{\rm S}}

\def    \rhoC		{\rho_{\rm C}}
\def    \Bsil		{B_{\rm S}}
\def    \Bgra		{B_{\rm C}}

\def    \stohd		{{\rm \left[Si/H\right]_{\rm dust}}}

\def    \ctohc		{{\rm \left[C/H\right]_{\rm cold}}}
\def    \ctohw		{{\rm \left[C/H\right]_{\rm warm}}}
\def    \xo             {x_{\rm o}}
\def \ameansil {{\langle a \rangle}_{\rm sil}}
\def \ameangra {{\langle a \rangle}_{\rm gra}}
\def \amean    {{\langle a \rangle}}
\countdef\decade=200
\advance\decade by \year
\countdef\hours=201
\advance\hours by \time
\divide\hours by 60
\countdef\mins=202
\advance\mins by \hours
\multiply\mins by 60
\multiply\hours by 100
\countdef\miltime=203
\advance\miltime by \hours
\advance\miltime by \time
\advance\miltime by -\mins
\def\today{\number\decade.\number\month.\number\day.\number\miltime}

\shorttitle{Silicon and the Ultraviolet Extinction}
\title{
\vspace*{-2.0em}
{\normalsize\rm Accepted for publication in
               {\it The Astrophysical Journal}}\\
\vspace*{1.0em}
Interstellar Silicon Depletion
and the Ultraviolet Extinction
\\{\small DRAFT: \today ~~}
}
\author{Ajay Mishra\altaffilmark{1,2}
        and Aigen Li\altaffilmark{1}}
\altaffiltext{1}{Department of Physics and Astronomy,
                 University of Missouri,
                 Columbia, MO 65211, USA;
                 {\sf lia@missouri.edu}
                 }
\altaffiltext{2}{Department of Science,
                 Technology and Mathematics, 
                 Lincoln University,
                 Jefferson City, MO 65101, USA;
                 {\sf 
                 mishraa@lincolnu.edu}
                 }

\begin{document}

\begin{abstract}
Spinning small silicate grains were recently
invoked to account for the Galactic foreground
anomalous microwave emission. These grains,
if present, will absorb starlight
in the far ultraviolet (UV). 
There is also renewed interest
in attributing the enigmatic 2175$\Angstrom$
interstellar extinction bump to small silicates.
To probe the role of silicon in the UV extinction,
we explore the relations between 
the amount of silicon required to 
be locked up in silicates $\sidust$
and the 2175$\Angstrom$ bump
or the far-UV extinction rise,
based on an analysis of the extinction
curves along 46 Galactic sightlines 
for which the gas-phase silicon abundance
$\sigas$ is known.
%
We derive $\sidust$ either from
$\siism-\sigas$ or from the 
Kramers-Kronig relation which relates
the wavelength-integrated extinction
to the total dust volume,
where $\siism$ is the interstellar silicon
reference abundance and taken to be
that of proto-Sun or B stars.
We also derive $\sidust$ from fitting
the observed extinction curves
with a mixture of amorphous silicates
and graphitic grains.
We find that in all three cases
$\sidust$ shows no correlation
with the 2175$\Angstrom$ bump,
while the carbon depletion $\cdust$
tends to correlate with the 2175$\Angstrom$ bump.
This supports carbon grains instead of silicates
as the possible carrier of the 2175$\Angstrom$ bump.
We also find that neither $\sidust$ nor $\cdust$
alone correlates with the far-UV extinction,
suggesting that the far-UV extinction is
a combined effect of small carbon grains
and silicates.
\end{abstract}


\section{Introduction}\label{sec:intro}
Silicon, an abundant metal element in the Universe, 
is highly depleted from the gas phase 
in the diffuse interstellar medium (ISM;
e.g., see Jenkins 1987, 2009)
as revealed by the weak ultraviolet (UV) 
absorption lines of Si\,II, the dominant form
of gas-phase silicon in the diffuse ISM  
(e.g., see van Steenberg \& Shull 1988). 
The silicon atoms missing from the gas phase
are thought to have been locked up in 
solid silicate dust grains (e.g., see Draine 1990). 
Silicate dust is ubiquitously seen in a wide variety of
astrophysical environments through the absorption
or emission spectral features arising from
the Si--O and O--Si--O vibrational modes
occurring respectively at 9.7 and 18$\mum$  
(see Henning 2010). In the diffuse ISM,
these features are seen in absorption
and their spectral profiles are smooth 
and lack fine structures, indicating a predominantly
amorphous composition (Li \& Draine 2001,
Kemper et al.\ 2004, Li et al.\ 2007).

It has been well recognized that silicate grains
are a major contributor to the interstellar extinction.
Assuming all Si, Mg, and Fe elements of solar abundances
(Asplund et al.\ 2009) are condensed in silicate dust  
with a stoichiometric composition of MgFeSiO$_4$ 
and a characteristic size of $a$\,$\approx$\,0.1$\mum$,
one can estimate the contribution of silicate dust to 
$\AV$, the extinction in the visual ($V$) band from
\begin{eqnarray}\label{eq:AV2NH}
\nonumber
\left(\frac{\AV}{N_{\rm H}}\right)_{\rm sil} & \approx &
   1.086\,\pi a^2 Q_{\rm ext}(V) N_{\rm sil}/N_{\rm H}\\
& \approx & \frac{1.086\,\pi a^2 Q_{\rm ext}(V)
\nonumber 
 \left(\sum_{\rm X=Si,Mg,Fe}\xsun\mux\,+\,4\,\sisun\muo\right)
 \muh}{\left(4/3\right)\pi a^3\rho_{\rm sil}}\\
& \approx & 3.25\times 10^{-22}\magni\cm^2~,
\end{eqnarray}
where $N_{\rm H}$ is the hydrogen column density,
$N_{\rm sil}$ is the column density of silicate dust,
$\rho_{\rm sil}$\,$\approx$3.5$\g\cm^{-3}$ is 
the mass density of silicate material, 
$Q_{\rm ext}(V)$ is the visual extinction efficiency 
of submicron-sized silicate dust which 
is taken to be $Q_{\rm ext}(V)$\,$\approx$1.5
(see Li 2009),
$\xsun$ is the solar abundance (relative to H)
of element X ($\sisun\approx32.4\pm2.2\ppm$,
$\mgsun\approx39.8\pm3.7\ppm$, and
$\fesun\approx31.6\pm2.9\ppm$, Asplund et al.\ 2009),
$\mux$ is the atomic weight of element X
($\mux\approx16,\,28,\,24,\,56$ 
for O, Si, Mg, and Fe, respectively), and 
$\muh\approx1.66\times10^{-24}\g$ is the mass of  
a hydrogen atom. Eq.\,\ref{eq:AV2NH} estimates
that silicate dust accounts for $\simali$60\% 
of the interstellar extinction for which
$\left(\AV/\NH\right)_{\rm obs}
\approx5.3\times10^{-22}\magni\cm^2$
(Bohlin et al.\ 1978).
Admittedly, this estimation is somewhat 
simplified since it is unlikely for interstellar 
silicate dust to have a single size of 
$a$\,=\,0.1$\mum$. Being highly
processed in the ISM by supernovae shocks
through sputtering and shattering
(Draine 2003), silicate dust is expected
to have a range of sizes. Without knowing
the silicate dust size distribution,
Mishra \& Li (2015) applied the Kramers-Kronig
(KK) relation of Purcell (1969) to estimate that 
silicate dust contributes to $\simali$48\%
of the total wavelength-integrated interstellar 
extinction which is obtained by integrating 
the observed interstellar extinction over 
wavelength from the far-UV 
to the far infrared (IR).
Indeed, all modern interstellar dust models
assume amorphous silicate to be a major
dust species (e.g., Mathis et al.\ 1977, 
Draine \& Lee 1984, Duley et al.\ 1989,
D\'esert et al.\ 1990, Siebenmorgen \& Kr\"ugel 1992, 
Mathis 1996, Li \& Greenberg 1997, 
Weingartner \& Draine 2001,
Li \& Draine 2001a, Zubko et al.\ 2004, 
Jones et al.\ 2013, Wang et al.\ 2015).   

In the Milky Way, the variation of 
the extinction $A_\lambda$
with wavelength $\lambda$, 
known as the extinction curve 
or the extinction law, is characterized by 
a nearly linear increase with $\lambda^{-1}$
in the near-IR, visible and near-UV,
a broad absorption bump at about
$\lambda^{-1}$\,$\approx$\,4.6$\mum^{-1}$
($\lambda$\,$\approx$\,2175$\Angstrom$),
and a steep rise into the far-UV
at $\lambda^{-1}$\,$\approx$\,10$\mum^{-1}$,
the shortest wavelength at which the dust extinction
has been measured (see Li et al.\ 2015).
Although a consensus has been achieved
that silicate dust is a major contributor of
the interstellar extinction,  it is not exactly clear
to what degree silicate dust is responsible for
the UV and far-UV extinction.

Very recently, Haris et al.\ (2016) determined
the gas-phase abundance of silicon ($\sigas$)
for 131 Galactic sightlines using archival data.
Assuming the interstellar abundance of silicon
to be solar (i.e., $\siism=\sisun$), they derived 
the silicon depletion (i.e., $\sidust=\siism-\sigas$)
for each sightline. Haris et al.\ (2016) further 
examined the UV extinction curves of 16 sightlines 
(of these 131 sightlines) and explored
the relation between the silicon depletion 
and the 2175$\Angstrom$ extinction bump
as well as the far-UV extinction rise.
Although the derived Pearson 
correlation coefficients
($R\approx-0.42$ for the silicon depletion
and the 2175$\Angstrom$ bump,
and $R\approx-0.32$ for the silicon depletion
and the far-UV rise) are by no means substantial,
they claimed that the silicon depletion 
positively ``correlates'' with 
the 2175$\Angstrom$ bump
and the far-UV extinction rise.
They further argued that these ``correlations'' 
imply that silicon plays a significant role 
in both the 2175$\Angstrom$ bump 
and the far-UV rise. 

The 2175$\Angstrom$ bump is the strongest 
spectroscopic extinction feature 
of the diffuse ISM (Draine 1989). 
Its carrier remains unidentified over half a century 
after its first detection (Stecher 1965). 
Although it is commonly attributed to 
small aromatic carbonaceous materials 
like nano-sized graphitic grains
(e.g., see Stecher \& Donn 1965, 
Draine \& Malhotra 1993, Mathis 1994)
or polycyclic aromatic hydrocarbon (PAH) molecules
(Joblin et al.\ 1992, Li \& Draine 2001a,
Cecchi-Pestellini et al.\ 2008, 
Steglich et al.\ 2010, Mulas et al.\ 2013,
Bekki et al.\ 2015),
Duley (1985) and Steel \& Duley (1987)
ascribed the 2175$\Angstrom$ bump
to small silicates or (Mg, Si) oxides.  

The 2175$\Angstrom$ bump
is an absorption feature 
with no scattered component,
with the detection of scattering 
reported only in two reflection nebulae 
(Witt et al.\ 1986). 
While the strength and width of
the 2175$\Angstrom$ bump
vary with environment, its peak wavelength 
is nearly invariant (see Draine 1989).
The nondetection of scattering and 
the stable peak wavelength imply that 
the carrier of the 2175$\Angstrom$ bump
is sufficiently small (i.e., nano-sized) 
to be in the Rayleigh limit (Mathis 1994).
If the 2175$\Angstrom$ bump is indeed
related to silicate grains as suggested by
Duley (1985), Steel \& Duley (1987) 
and Haris et al.\ (2016), there would 
exist a population of nano-sized 
silicate grains in the diffuse ISM. 

The far-UV extinction is also predominantly 
absorptive. General results concerning scattering
by small particles indicate that the far-UV 
extinction arises from nano-sized grains
(see Draine 1995), although the exact sizes
of these grains can not be constrained by
the far-UV extinction (see Wang et al.\ 2015).   
Therefore, the correlation between the silicon 
depletion and the far-UV extinction rise derived 
by Haris et al.\ (2016) also implies the existence
of an appreciable number of nano silicate grains
in the diffuse ISM.

Very recently, nano silicate grains are of 
renewed interest. 
Hoang et al.\ (2016) and Hensley \& Draine (2017a)
argued that the so-called ``anomalous microwave 
emission'' (AME), an important Galactic foreground 
of the cosmic microwave background radiation
in the $\simali$10--100\,GHz region,
could arise from spinning nano-sized silicate grains.
However, silicate nanoparticles will undergo 
single-photon heating in the ISM and emit at 
the 9.7$\mum$ Si--O feature. 
The nondetection of the 9.7$\mum$ 
emission feature in the diffuse ISM
allows one to place an upper limit
on the number of nano-sized interstellar
silicate grains (see Li \& Draine 2001b).

In light of these contradicting results
or hypothesis (i.e., whether the 2175$\Angstrom$
extinction bump is due to small silicate grains
or small carbonaceous grains, and
whether nano-sized silicate grains are
abundant in the diffuse ISM),
we are motivated to explore the role 
of silicate grains in the interstellar UV
extinction, with special attention paid
to the 2175$\Angstrom$ extinction bump
and the far-UV rise. 
To achieve this, we consider 
a larger sample of 46 sightlines 
for which both the UV extinction curves
and the gas-phase $\sigas$ abundances have
been observationally determined 
(see \S\ref{sec:sample}).
In contrast, Harris et al.\ (2016)
only considered 16 sightlines.
We first investigate the relation 
between the silicon depletion $\sidust$ 
and the UV extinction by assuming an interstellar 
reference abundance\footnote{%
   The reference abundance 
   (also known as ``interstellar abundance'',  
   or ``cosmic abundance'') of an element
   is the total abundance of this element 
   (both in gas and in dust).
   }
of $\siism$ and deriving $\sidust$ from subtracting 
from $\siism$ off the gas-phase abundance 
$\sigas$ (see \S\ref{sec:depletion}).
We then examine the relation between $\sidust$ 
and the UV extinction by deriving $\sidust$ from
the KK relation of Purcell (1969) 
which relates the wavelength-integrated extinction 
to the total dust volume (see \S\ref{sec:kk}).
Alternatively, we will also derive $\sidust$ 
by modeling the observed extinction curve 
of each sightline in terms of 
the silicate-graphite model (see \S\ref{sec:extmod})
and then compare the derived $\sidust$
with the observed UV extinction.
Finally, the results are discussed in \S\ref{sec:discussion}
and the major conclusions are summarized 
in \S\ref{sec:conclusion}.
\section{The Sample}\label{sec:sample}
We compile from the literature all the Galactic
sightlines for which both the UV extinction curves
and the silicon gas-phase abundances $\sigas$
have been observationally determined. 
Resultantly, we arrive at a sample of 46 sightlines 
(see Table~\ref{tab:kk})
which is a subsample of the 131 sightlines 
of Haris et al.\ (2016). 
For each sightline, we take the $\sigas$ abundance
from Haris et al.\ (2016) and the extinction parameters
$c_j^{\prime}$ ($j$\,=\,1, 2, 3, 4), $\xo$ and $\gamma$
(see below) from Jenniskens \& Greenberg (1993), 
Valencic et al.\ (2004), Lewis et al.\ (2005), 
and Gordon et al.\ (2009).   

Following Valencic et al.\ (2004)
and Gordon et al.\ (2009),
we ``construct'' the UV extinction curve 
at $3.3 < \lambda^{-1} < 8.7\mum^{-1}$
of each sightline, as a function of
$x\equiv 1/\lambda$ (in $\mu$m$^{-1}$), 
the inverse wavelength, 
from the following formula:
\begin{equation}\label{eq:A2AV}
A_\lambda/A_V = c_1^{\prime} + c_2^{\prime}\,x 
              + c_3^{\prime}\,D(x,\gamma,\xo) 
              + c_4^{\prime}\,F(x) ~~~.
\end{equation}
This analytical formula consists of 
(i) a linear background term described by 
$c_1^{\prime}$ and $c_2^{\prime}$,
(ii) a Drude-profile term for the 2175$\Angstrom$ 
extinction bump (of peak $\xo$ and FWHM $\gamma$)
approximated as 
\begin{equation}
D(x,\gamma,\xo) \equiv \frac{x^2}
  {\left(x^2-\xo^2\right)^2 + x^2\gamma^2} ~~~,
\end{equation}
and (iii) a far-UV nonlinear-rise term
at $\lambda^{-1} > 5.9\mum^{-1}$
represented by
\begin{equation}
F(x) = \left\{\begin{array}{lr} 
0 ~, & x < 5.9\mum^{-1} ~~~,\\

0.5392\,\left(x-5.9\right)^2 
     + 0.05644\,\left(x-5.9\right)^3 ~, 
 & x \ge 5.9\mum^{-1} ~~~.\\
\end{array}\right.
\end{equation}
This parametrization,
in which $c_3^{\prime}$ and $c_4^{\prime}$ respectively define 
the strength of the 2175$\Angstrom$ extinction bump 
and the strength of the nonlinear far-UV rise,
was originally introduced
by Fitzpatrick \& Massa (1990; hereafter FM90)
for the interstellar reddening
\begin{equation}
E(\lambda-V)/E(B-V) = R_V \left(A_\lambda/A_V - 1\right)
              = c_1 + c_2\,x 
              + c_3\,D(x,\gamma,\xo) 
              + c_4\,F(x) ~~~,
\end{equation}
where $E(\lambda-V) \equiv A_\lambda - A_V$,
$E(B-V) \equiv A_B - A_V$, 
$A_B$ is the $B$-band extinction,
and $R_V\equiv A_V/E(B-V)$ is
the optical total-to-selective extinction ratio.
The $c_j$ parameters of FM90 are related
to that of Valencic et al.\ (2004) and 
Gordon et al.\ (2009), $c_j^{\prime}$, through  
\begin{equation}
c_j^{\prime} = \left\{\begin{array}{lr} 
c_j/R_V + 1 ~, & j=1 ~~~,\\
c_j/R_V ~, & j=2, 3, 4 ~~~.\\
\end{array}\right.
\end{equation}
Finally, for each sightline
we compute the optical/near-IR extinction 
at $0.3 < \lambda^{-1} < 3.3\mum^{-1}$
from the $\RV$-based CCM parametrization 
(see Cardelli et al.\ 1989).
We then smoothly join the UV extinction
at $\lambda^{-1} > 3.3\mum^{-1}$
to the optical/near-IR extinction
at $\lambda^{-1} < 3.3\mum^{-1}$.
In Figures~\ref{fig:extmod1}--\ref{fig:extmod5} 
we show the UV/optical/near-IR extinction curves 
of 39 sightlines constructed as above.
The extinction curves of the other seven sightlines 
have already been constructed and modeled previously
(see Mishra \& Li 2015).
\section{Silicon Depletion Inferred from 
           $\siism$ and $\sigas$}\label{sec:depletion}
Elements in the ISM exist in the form of gas or dust.
The interstellar gas-phase abundances of elements
can be measured from their optical and UV
spectroscopic absorption lines. 
The elements ``missing'' from the gas phase 
are bound up in dust grains,
known as ``interstellar depletion''. 
The dust-phase abundance of an element is 
derived by assuming a reference abundance
and then from which subtracting off 
the gas-phase abundance. 

Historically, the interstellar abundances of
the dust-forming elements C, O, Mg, Si, Fe 
were commonly assumed to be solar. 
However, Lodders (2003) argued that the currently 
observed solar photospheric abundances (relative to H) 
must be lower than those of the proto-Sun 
because helium and other heavy elements 
may have settled toward the Sun's 
interior since the time of its formation 
$\simali$4.55$\Gyr$ ago.
Lodders (2003) further suggested that 
the protosolar abundances 
derived from the combined considerations of 
the present-day photospheric abundances 
and all the possible settling effects 
are more representative of 
the true interstellar abundances. 
On the other hand, it has also been argued that 
the interstellar abundances,
because of their young ages,
might be better represented by 
those of B stars and young F, G stars 
(e.g., see Snow \& Witt 1995, 1996, 
Sofia \& Meyer 2001).
Therefore, in the following we will consider 
two sets of reference abundances 
for the dust-forming element:
$\xism =\xsun$ --- the protosolar abundances 
of Lodders (2003),
and $\xism =\xstar$ --- the B-star abundances 
of Przybilla et al.\ (2008). 

With $\siism=\sisun\approx40.7\pm1.9\ppm$ 
of Lodders (2003) 
or $\siism=\sistar\approx31.6\pm1.5\ppm$ 
of Przybilla et al.\ (2008), we respectively derive
the silicon depletion $\sidust=\siism-\sigas$
for each sightline and then compare with 
the strength of the 2175$\Angstrom$ extinction bump
(measured by $c_3^{\prime}$) 
and the strength of the nonlinear far-UV rise
(measured by $c_4^{\prime}$). 
As shown in Figure~\ref{fig:c3_si2h_ism},
neither the 2175$\Angstrom$ bump
nor the far-UV rise correlates 
with the silicon depletion.\footnote{%
   Although somewhat arbitrary, 
   we suggest that, for two variables
   to be considered to be (even weakly) correlated, 
   the Pearson correlation coefficient ($R$) 
   should at least exceed 0.5
   (e.g., see {\sf https://explorable.com/statistical-correlation}).
   }
This is true for both sets of
reference-abundances:
the exact value of the interstellar
silicon abundance is unimportant as long 
as we assume that it is the same for all
directions.
The correlation plots are identical
for the two sets of reference-abundances
except for a shift in ordinate.

\section{Silicon Depletion Inferred from 
            the Kramers-Kronig Relation\label{sec:kk}}
The approach discussed in \S\ref{sec:depletion} 
requires the knowledge of $\siism$ and $\sigas$.
We note that $\sigas$ is often difficult to measure
since silicon is often highly depleted and the Si\,II 
absorption lines are rather weak. In this section we take
an alternative, carbon-based approach which makes use
of the KK relation of Purcell (1969)
and does not require the knowledge of $\siism$ and $\sigas$.

Let $\cism$ be the total interstellar carbon abundance 
(relative to H), and $\cgas$ be the gas-phase carbon 
abundance of a given sightline. 
We derive the total volume (per H nucleon) 
of carbon dust for this sightline from
\begin{equation}
 \frac{\VC}{\rm H} = 
\left\{\cism-\cgas\right\}\times 12\,\mH/\rho_{\rm C} ~~,
\end{equation}
where $\rhoC$ is the mass density of the carbon dust 
($\rhoC\approx2.24\g\cm^{-3}$ for graphite).
We can apply the KK relation of Purcell (1969) 
to gain insight into the amount of extinction
resulting from such an amount of carbon dust;
particularly, the wavelength-integrated extinction
is directly related to the dust volume through
\begin{equation}\label{eq:kk}
\int_{0}^{\infty} \left(\frac{A_\lambda}{\NH}\right)_{\rm C} d\lambda 
= 1.086\times 3 \pi^2 \FC \frac{\VC}{\rm H} ~~,
\end{equation}
where $\left(A_\lambda/\NH\right)_{\rm C}$ is 
the extinction (per H column) caused by carbon dust,
and $\FC$ is a dimensionless factor which depends 
only upon the grain shape and the static (zero-frequency) 
dielectric constant $\varepsilon_0$ of the grain material.
For conducting, graphitic grains 
of moderately elongated shapes,
Mishra \& Li (2015) derived $\FC\approx1.25$ 
(see their Figure~1).

Let $\left(A_\lambda/\NH\right)_{\rm obs}$ be
the observed extinction (per H column) of
a given sightline. If we assume that there are
two major dust populations in the ISM
--- amorphous silicate and carbon dust,
the extinction contributed by silicate dust, 
$\left(A_\lambda/\NH\right)_{\rm sil}$,
can be obtained from
\begin{equation}\label{eq:Asil}
\int_{0}^{\infty} \left(\frac{A_\lambda}{\NH}\right)_{\rm sil} d\lambda 
= \int_{0}^{\infty} \left(\frac{A_\lambda}{\NH}\right)_{\rm obs} d\lambda 
- \int_{0}^{\infty} \left(\frac{A_\lambda}{\NH}\right)_{\rm C} d\lambda ~~.
\end{equation}
To account for such an amount of extinction,
one requires a total silicate dust volume of
\begin{equation}\label{eq:Vsil1}
\frac{\Vsil}{\rm H}  = 
\frac
{\int_{0}^{\infty} \left(A_\lambda/\NH\right)_{\rm sil} d\lambda} 
{1.086\times 3 \pi^2 \Fsil} ~~,
\end{equation}
where $\Fsil$, like $\FC$, is a dimensionless factor.
For moderately elongated silicate grains,
$\Fsil\approx0.7$ (see Figure~1 in Mishra \& Li 2015).
We then derive the silicon depletion 
from the silicate dust volume
\beq\label{eq:Si2H}
\sidust = \frac
{\left(\Vsil/\rmH\right)\rho_{\rm sil}}
{\mu_{\rm sil}\mH} ~~,
\eeq
where $\mu_{\rm sil}=172$ is the molecular weight 
for MgFeSiO$_4$. 
Therefore, we can infer $\sidust$ from $\cism$, $\cgas$
and $\left(A_\lambda/\NH\right)_{\rm obs}$,
without knowing $\siism$ and $\sigas$.
Also, we do not need to know the exact
dust properties (e.g., sizes, compositions) 
except for an assumed dust species 
we need to specify its mass density ($\rho$), 
molecular weight ($\mu$) 
and whether the dust material is conducting 
(i.e., $\varepsilon_0\rightarrow\infty$) 
or dielectric (i.e., $\varepsilon_0$ is finite).

Observationally, for many of our sightlines
considered in this work
the extinction per H nucleon is only known 
over a limited range of wavelengths
(e.g., $0.3 < \lambda^{-1} < 8.7\mum^{-1}$).
We approximate $\int_{0}^{\infty} 
\left(A_\lambda/\NH\right)_{\rm obs}\,d\lambda$
by $\int_{912\Angstrom}^{1000\mum} 
\left(A_\lambda/\NH\right)_{\rm obs}\,d\lambda$.\footnote{%
  The silicon depletion derived from 
  $\int_{912\Angstrom}^{1000\mum} 
  \left(A_\lambda/\NH\right)_{\rm obs}\,d\lambda$
   is a lower limit 
   since $\int_{912\Angstrom}^{1000\mum} 
   \left(A_\lambda/\NH\right)_{\rm obs}\,d\lambda
   < \int_{0}^{\infty} 
   \left(A_\lambda/\NH\right)_{\rm obs}\,d\lambda$.
   }
We first use eq.\,\ref{eq:A2AV}
to extrapolate the UV extinction 
at $\lambda^{-1} < 8.7\mum^{-1}$ 
to $\lambda=912\Angstrom$.\footnote{%
  Gordon et al.\ (2009) studied the extinction curves
  of 75 Galactic sightlines 
  obtained with 
  the {\it Far Ultraviolet Spectroscopic Explorer} 
  (FUSE) at $905 < \lambda < 1187\Angstrom$ 
  and the {\it International Ultraviolet Explorer} (IUE) 
  at $1150 < \lambda < 3300\Angstrom$.  
  They found that the extrapolation
  of the UV extinction at $3.3 < \lambda^{-1} < 8.7\mum^{-1}$
  obtained by IUE
  is generally consistent with the far-UV extinction 
  at $8.4 < \lambda^{-1} < 11\mum^{-1}$
  obtained by FUSE.
  }
For $0.001 < \lambda^{-1} < 0.3\mum^{-1}$,
we adopt the model $A_\lambda/\NH$ values of
Weingartner \& Draine (2001; WD01) 
for the $R_V=3.1$ diffuse ISM
(see Figure~16 in Li \& Draine 2001).
This is justified since the near- and mid-IR
extinction at $\lambda > 0.9\mum$ does not
seem to vary much among different environments
(see Wang et al.\ 2013, 2014).
In Table~\ref{tab:kk} we tabulate 
the wavelength-integrated extinction for each sightline.

For the interstellar carbon reference abundance,
similar to silicon in \S\ref{sec:depletion}, 
we also consider 
two sets of reference abundances: 
$\cism =\csun\approx288\pm27\ppm$ 
--- the protosolar abundance
of Lodders (2003),
and $\cism =\cstar\approx214\pm20\ppm$
--- the B-star abundance of Przybilla et al.\ (2008). 
The gas-phase $\cgas$ abundance are not known
for all sightlines. We estimate $\cgas$ from 
the hydrogen number density $\nH$,
using the four-parameter Boltzmann function
originally proposed by Jenkins et al.\ (1986) 
and subsequently modified by 
Cartledge et al.\ (2004, 2006):
\begin{equation}\label{eq:cgas}
\cgas = \ctohc+\frac{\ctohw-\ctohc}
{1+\exp\left\{\log_{10}\left(\nH/n_0\right)/m\right\}}~~,
\end{equation}
where $\ctohw$ and $\ctohc$ are respectively 
the carbon gas-phase abundance levels 
for low and high mean sightline densities,
$n_{0}$ is a parameter with a dimension of 
hydrogen number density, 
and $m$ is a dimensionless parameter.\footnote{%
   We note that there was a typo in 
   the expressions of Cartledge et al.\ (2004, 2006):
   the term $\left(\nH-n_0\right)$ in eq.\,1
   of  Cartledge et al.\ (2004)
   and eq.\,1 of Cartledge et al.\ (2006)
   should actually be 
   $\left(\log_{10}\nH-\log_{10} n_0\right)$.
   }
As shown in Figure~\ref{fig:c2h_gas},
with $\ctohw\approx480\pm48\ppm$,
$\ctohc\approx100.34\pm14.63\ppm$, 
$\log_{10}\left( n_{0}/{\rm cm^{-3}}\right)
\approx-0.919\pm-0.103$, 
and $m\approx0.33\pm0.12$,
the Boltzmann-like function fits reasonably
well the $\nh$--$\cgas$ relation for those
sightlines of which both $\nh$ and $\cgas$ are known.
For those sightlines of which $\nH$ 
(but not $\cgas$) has been observationally determined,
we estimate the gas-phase $\cgas$ abundance
from the hydrogen number density $\nH$ and 
list in Table~\ref{tab:modpara}.

With $\left(A_\lambda/\NH\right)_{\rm obs}$ constructed
as above and $\cgas$ estimated from $\nH$, 
we now use the KK relation of Purcell (1969)
to derive the silicon depletions $\sidust$ 
for all 46 sightlines and then compare $\sidust$
with the UV extinction. 
As shown in Figure~\ref{fig:c3_si2h_kk},
$\sidust$ does not correlate with $c_3^{\prime}$
(i.e., the 2175$\Angstrom$ extinction bump),
and nor does it correlate with $c_4^{\prime}$
(i.e., the far-UV extinction rise).
This is true no matter 
whichever is adopted as 
the interstellar reference abundance
--- the protosolar C/H abundance
or the B-star C/H abundance.
As already mentioned in \S\ref{sec:depletion},
the exact value of the interstellar
silicon abundance does not affect
the correlation coefficient but 
causes a shift in ordinate.

So far, for the carbon dust species we are
confined to graphite. However, this is actually
unnecessary. After all, the KK approach does not
really involve the optical properties and 
the size distribution of graphite.
All we have used are the mass density
of graphite ($\rho_{\rm gra}\approx2.24\g\cm^{-3}$) 
and the dimensionless factor $\FC$.
If we consider amorphous carbon instead of 
graphite, the same conclusion will be drawn
except the available carbon dust volume will
be increased by a factor of 
$\rho_{\rm gra}/\rho_{\rm ac}$,
where $\rho_{\rm ac}\approx1.8\g\cm^{-3}$ 
is the mass density of amorphous carbon.
As illustrated in Figure~1 of Mishra \& Li (2015),
for moderately elongated grains,
the dimensionless $\FC$ factor of amorphous carbon
($\FC\approx1.20$) 
differs only by $\simali$5\% from that of graphite
($\FC\approx1.25$).

\section{Silicon Depletion Inferred from 
            Interstellar Extinction Modeling}
               \label{sec:extmod}
The silicon depletion $\sidust$ derived from
the KK relation of Purcell (1969) in \S\ref{sec:kk}
is independent of any exact dust models 
except we just need to assume that 
the observed extinction is caused by
silicate dust and carbon dust and specify 
the mass densities of the relevant dust materials
and the static dielectric constant $\varepsilon_0$
which, together with the grain shape,
determines the dimensionless factor 
$\FC$ or $\Fsil$.

We now derive the silicon depletion
from fitting the observed extinction curve 
for each sightline.
We consider the silicate-graphite interstellar
grain model which consists of two separate dust
components: amorphous silicate and graphite
(Mathis et al.\ 1977, Draine \& Lee 1984).
We adopt an exponentially-cutoff power-law size
distribution for both components:
$dn_i/da = \nH B_i a^{-\alpha_i} \exp\left(-a/a_{c,i}\right)$
for the size range of
$50\Angstrom < a < 2.5\um$,
where $a$ is the spherical radius of the dust,
$dn_i$ is the number density of dust of type $i$
with radii in the interval [$a$, $a$\,$+$\,$da$],
$\alpha_i$ and $a_{c,i}$ are respectively
the power index and exponential cutoff size
for dust of type $i$, and
$B_i$ is the constant related to
the total amount of dust of type $i$.
The total extinction per H column
at wavelength $\lambda$ is given by
\begin{equation}\label{eq:Amod}
A_\lambda/\NH = 1.086
            \sum_i \int da \frac{1}{\nH} 
            \frac{dn_i}{da}
            C_{{\rm ext},i}(a,\lambda),
\end{equation}
where the summation is over the two grain types
(i.e., silicate and graphite),
$\NH$ is the hydrogen column density,
and $C_{{\rm ext},i}(a,\lambda)$
is the extinction cross section of
grain type $i$ of size $a$
at wavelength $\lambda$
which can be calculated 
from Mie theory (Bohren \& Huffman 1983)
using the dielectric functions of
``astronomical'' silicate and graphite 
of Draine \& Lee (1984).

In fitting the extinction curve,
for a given sightline, 
we have six parameters:
the size distribution power indices 
$\alphaS$ and $\alphaC$ 
for silicate and graphite, respectively;
the exponential cutoff sizes 
$\acS$ and $\acC$, respectively;
and $\Bsil$ and $\Bgra$.
We derive the silicon and carbon depletions from
\begin{equation}\label{eq:Si2H}
\sidust = \left(\nH\Bsil/172\mH\right) 
        \int da \left(4\pi/3\right) a^3\,\rhosil
               a^{-\alphaS}
               \exp\left(-a/\acS\right) ~~,
\end{equation}
\begin{equation}\label{eq:C2Ha}
\cdust = \left(\nH\Bgra/12\mH\right) 
        \int da \left(4\pi/3\right) a^3\,\rhogra
               a^{-\alphaC}
               \exp\left(-a/\acC\right) ~~,
\end{equation}
where we assume a stoichiometric composition of
MgFeSiO$_4$ for amorphous silicate.

For a given sightline,
we seek the best fit to the extinction 
between $0.3\mum^{-1}$ and $8\mum^{-1}$
by varying the size distribution power indices 
$\alphaS$ and $\alphaC$, and
the upper cutoff size parameters 
$\acS$ and $\acC$.
Following WD01,
we evaluate the extinction at 
100 wavelengths $\lambda_i$, 
equally spaced in $\ln\lambda$.
We use the Levenberg-Marquardt method
(Press et al.\ 1992) to minimize $\chi^2$ 
which gives the error in the extinction fit:
\begin{equation}
\chi^2 = \sum_i \frac{\left( \ln A_{\rm obs} 
       - \ln A_{\rm mod} \right)^2} {\sigma_i^2}~~~,
\end{equation}
where $A_{\rm obs}(\lambda_i)$ 
is the observed extinction at wavelength $\lambda_i$,
$A_{\rm mod}(\lambda_i)$ 
is the extinction computed
for the model at wavelength $\lambda_i$
(see eq.\,\ref{eq:Amod}), 
and the $\sigma_i$ are weights.  
Following WD01,
we take the weights $\sigma_i^{-1} = 1$ 
for $1.1 < \lambda^{-1} < 8\mum^{-1}$ 
and $\sigma_i^{-1} = 1/3$ for 
$\lambda^{-1} < 1.1\mum^{-1}$.

\begin{figure*}
\centering	
\vspace{-5mm}
\includegraphics[width=0.80\textwidth]{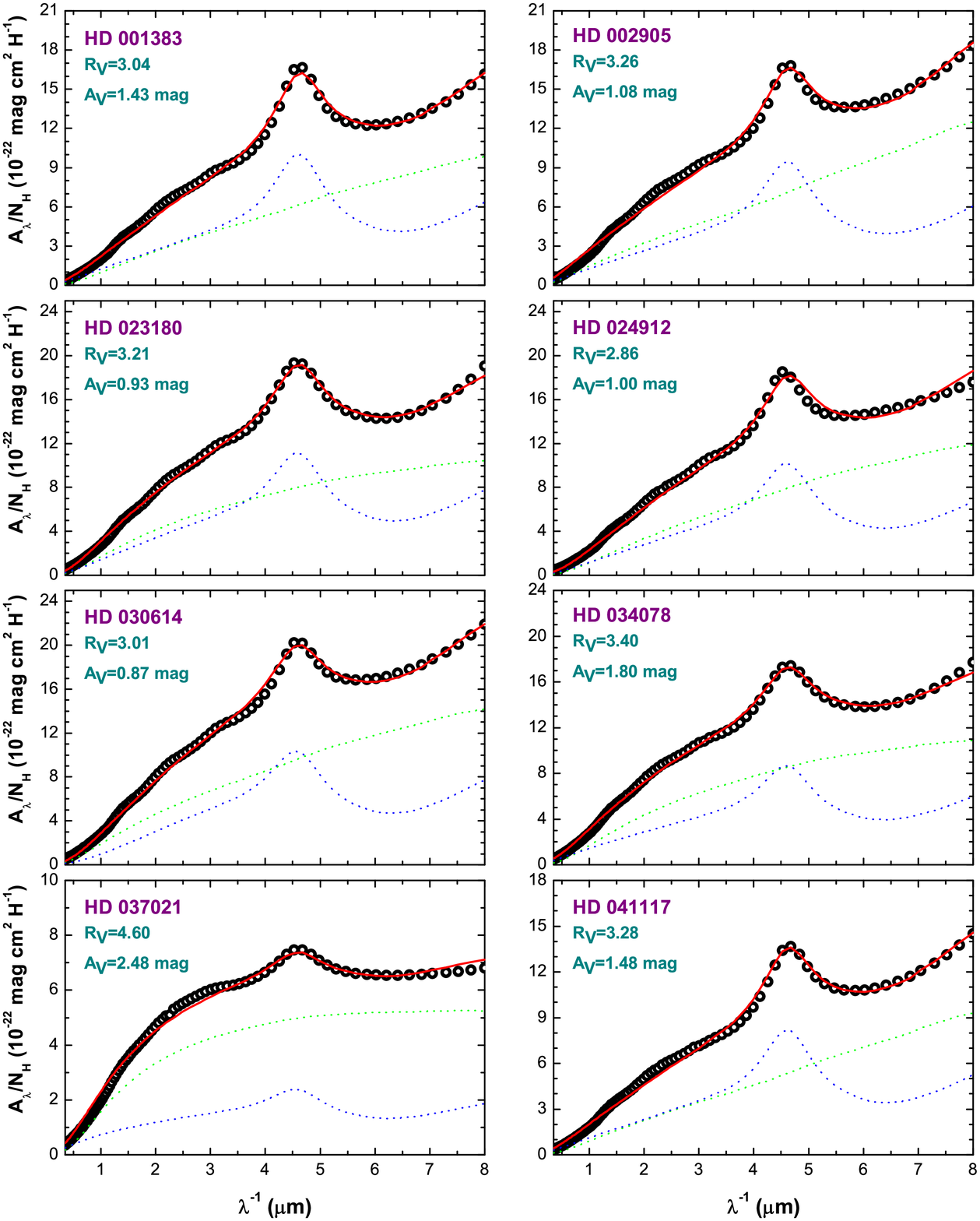}
\vspace{-2mm}
\caption{\footnotesize
         \label{fig:extmod1}
         Observed and model extinction curves of 
         HD\,001383, HD\,002905, HD\,023180,
         HD\,024912, HD\,030614, HD\,034078,
         HD\,037021, and HD\,04117.
         The observed extinction curves 
         are represented by 
         the FM90 parametrization 
         at $\lambda^{-1} > 3.3\mum^{-1}$
         and by the CCM parametrization 
         at $\lambda^{-1} < 3.3\mum^{-1}$
         (open black circle).
         The solid red line plots the model
         extinction curve which is a combination
         of amorphous silicate (dotted green line)
         and graphite (dashed blue line).
         }
\end{figure*}

\begin{figure*}
\centering	
\vspace{-5mm}
\includegraphics[width=0.80\textwidth]{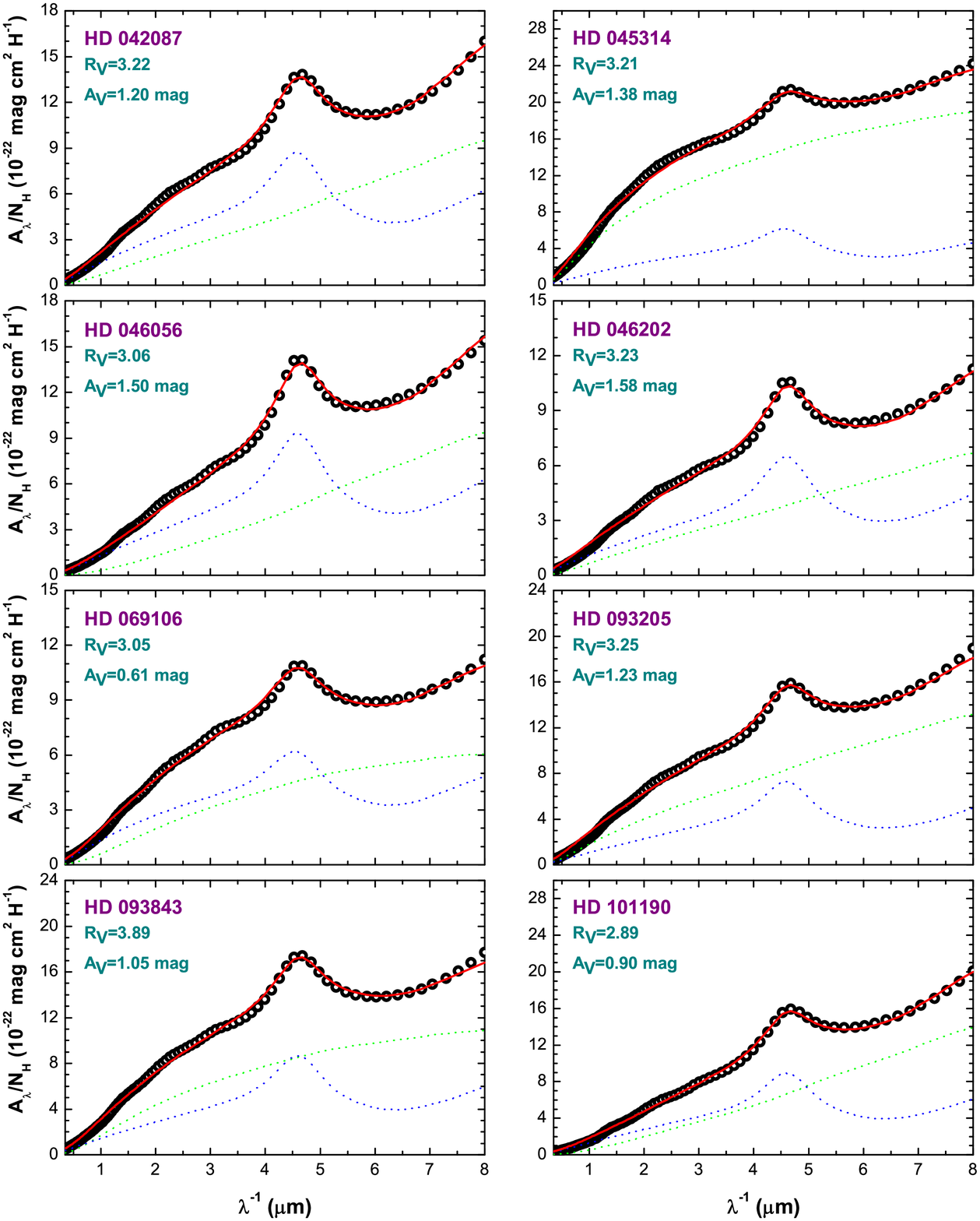}
\vspace{-2mm}
\caption{\footnotesize
        \label{fig:extmod2}
         Same as Figure~\ref{fig:extmod1}
         but for HD\,042087, HD\,045314, HD\,046056,
         HD\,046202, HD\,069106, HD\,093205,
         HD\,093843, and HD\,101190.
         }
\end{figure*}

\begin{figure*}
\centering	
\vspace{-5mm}
\includegraphics[width=0.80\textwidth]{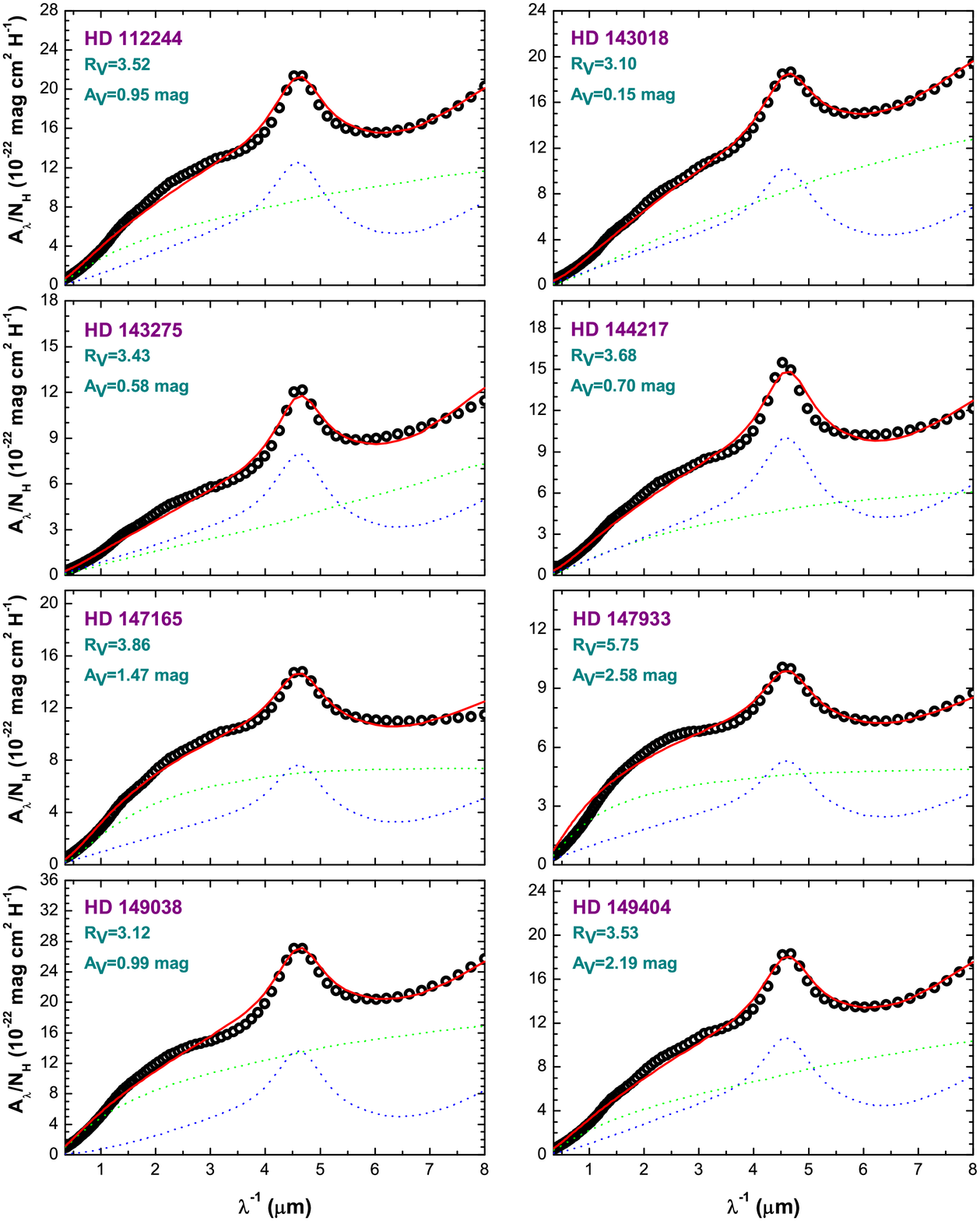}
\vspace{-2mm}
\caption{\footnotesize
        \label{fig:extmod3}
         Same as Figure~\ref{fig:extmod1}
         but for HD\,112244, HD\,143018,
         HD\,143275, HD\,144217, HD\,147165,
         HD\,147933, HD\,149038, and HD\,149404.
         }
\end{figure*}

\begin{figure*}
\centering	
\vspace{-5mm}
\includegraphics[width=0.80\textwidth]{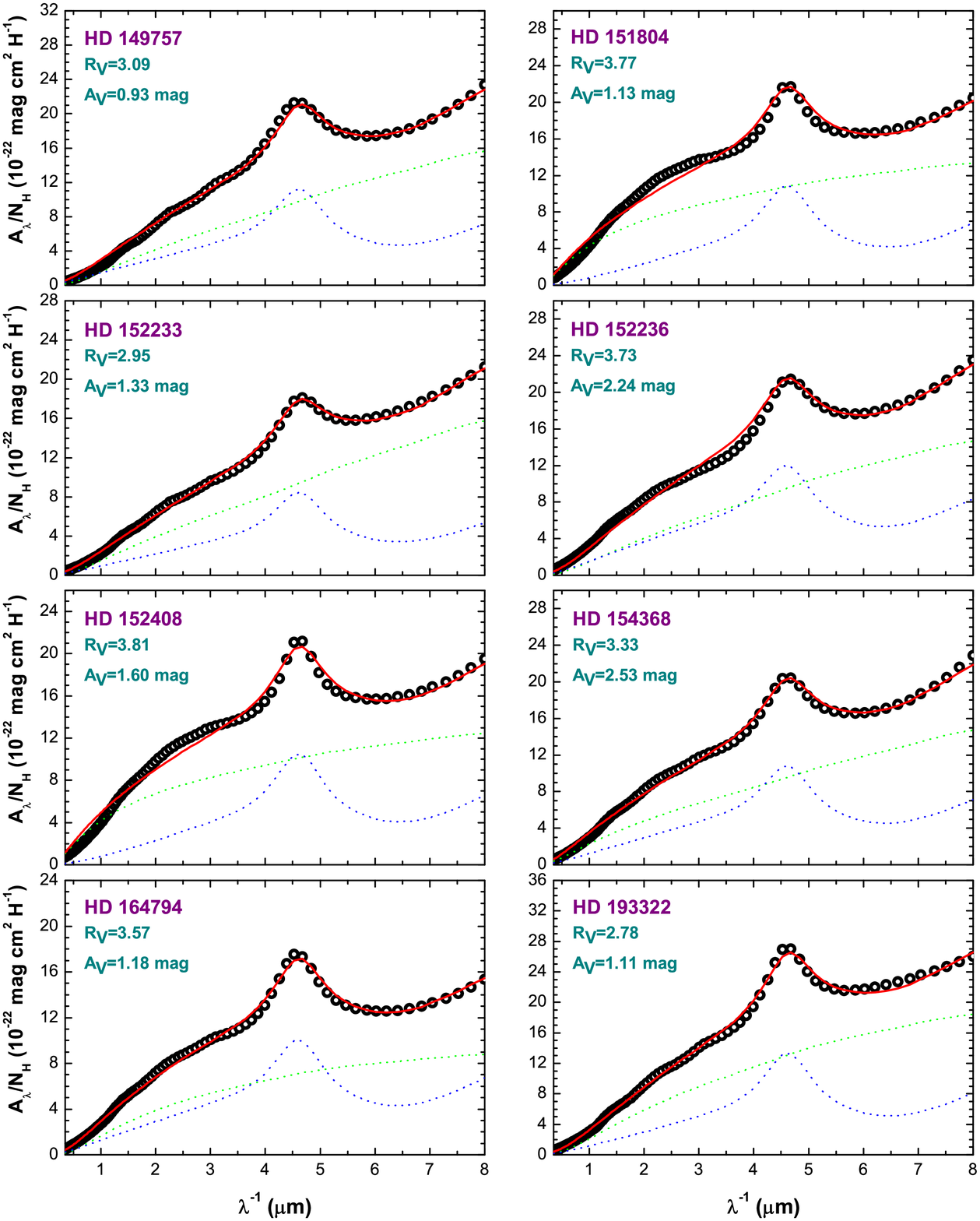}
\vspace{-2mm}
\caption{\footnotesize
        \label{fig:extmod4}
         Same as Figure~\ref{fig:extmod1}
         but for HD\,149757, HD\,151804,
         HD\,152233, HD\,152236, HD\,152408,
         HD\,154368, HD\,164794, and HD\,193322.
         }
\end{figure*}

\begin{figure*}
\centering	
\vspace{-5mm}
\includegraphics[width=0.80\textwidth]{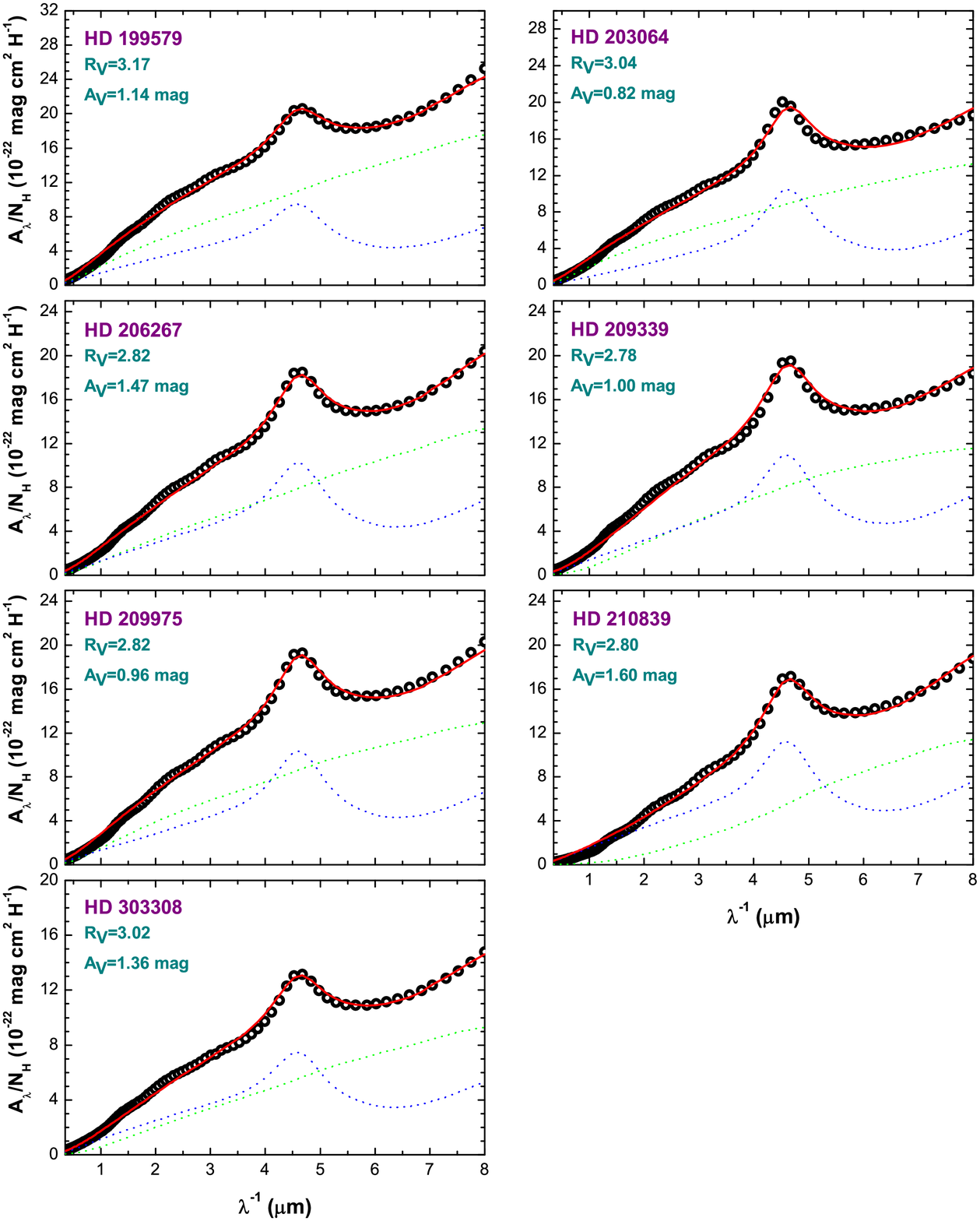}
\vspace{-2mm}
\caption{\footnotesize
        \label{fig:extmod5}
         Same as Figure~\ref{fig:extmod1}
         but for HD\,199579, HD\,203064,
         HD\,206267, HD\,209339, HD\,209975,
         HD\,210839, and HD\,303308.
         }
\end{figure*}

\begin{figure*}
\centering
\begin{tabular}{c}
\includegraphics[width=1.0\textwidth]{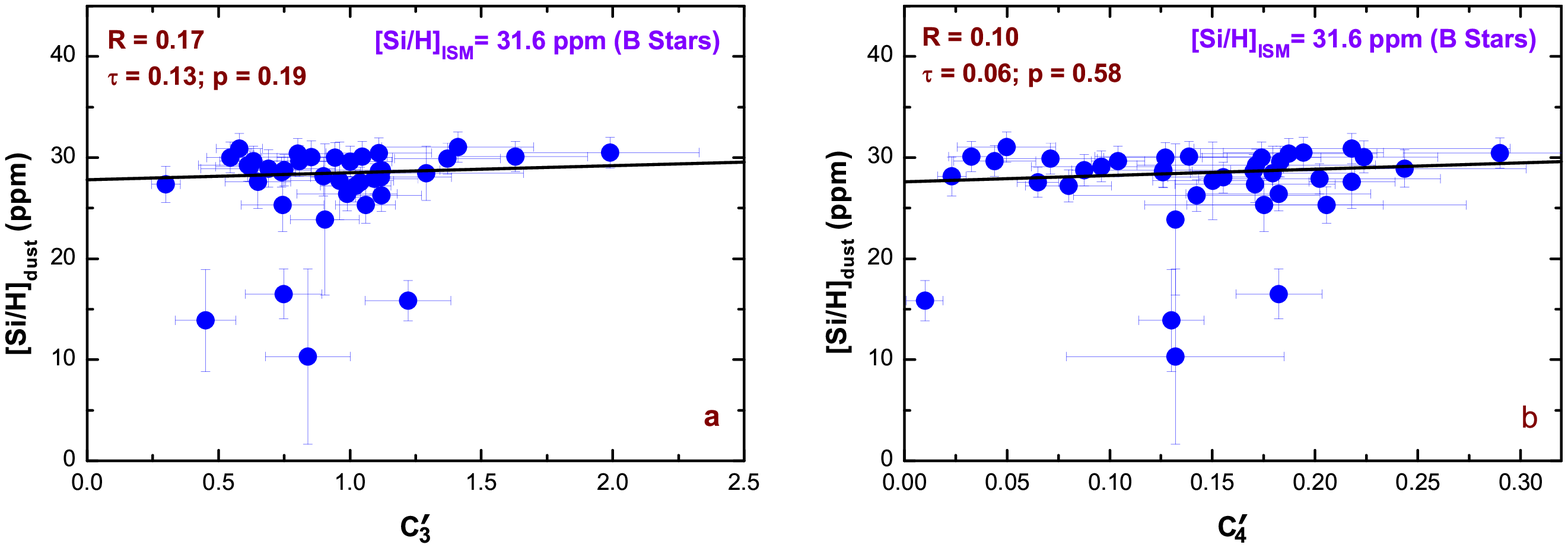}
\end{tabular}
\caption{\footnotesize
         \label{fig:c3_si2h_ism}
         Correlation diagrams 
         between the silicon depletion $\sidust$ 
         and the strength of the 2175$\Angstrom$ 
         extinction bump ($c_3^{\prime}$; left panel) or
	 the far-UV extinction rise ($c_4^{\prime}$; 
	 right panel), where $\sidust = \siism - \sigas$. 
         The interstellar silicon reference 
         abundance $\siism$ is assumed 
         to be that of B stars (Przybilla et al.\ 2008).
         The gas-phase silicon abundance $\sigas$
         is taken from Haris et al.\ (2016).
         Note that the exact value of the assumed 
         interstellar silicon reference abundance 
         (of proto-Sun or B stars)
         does not affect
         the correlation coefficient
         but the intercept.
         Also labelled are the Kendall's $\tau$
	 coefficient and the significance level $p$.
         }
\end{figure*}

\begin{figure*}
\centering
\includegraphics[width=0.5\textwidth]{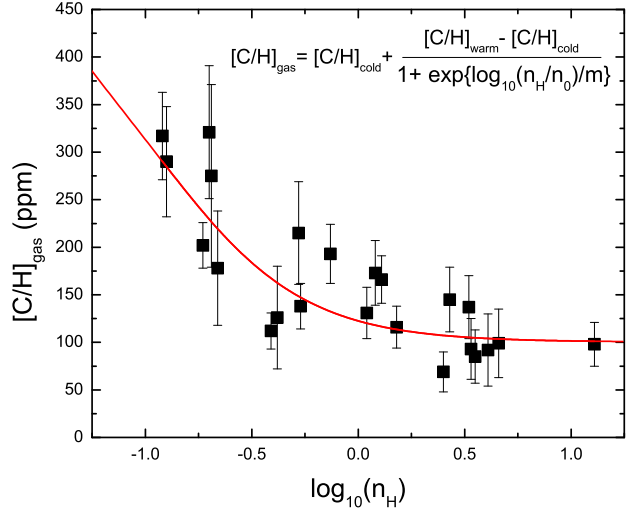}
\caption{\footnotesize
         \label{fig:c2h_gas}
         Density-dependent gas-phase carbon abundance 
         $\cgas$ (black squares) 
         fitted by a Boltzmann function (red line)
         characterized by a set of four parameters:
         $\ctohw\approx480\pm48\ppm$,
         $\ctohc\approx100\pm15\ppm$,
         $\log_{10}\left(n_0/{\rm cm^{-3}}\right)\approx-0.92\pm-0.10$, 
         and $m\approx0.33\pm0.12$.
         }
\end{figure*}

\begin{figure*}
\centering
\begin{tabular}{c}
\includegraphics[width=1.0\textwidth]{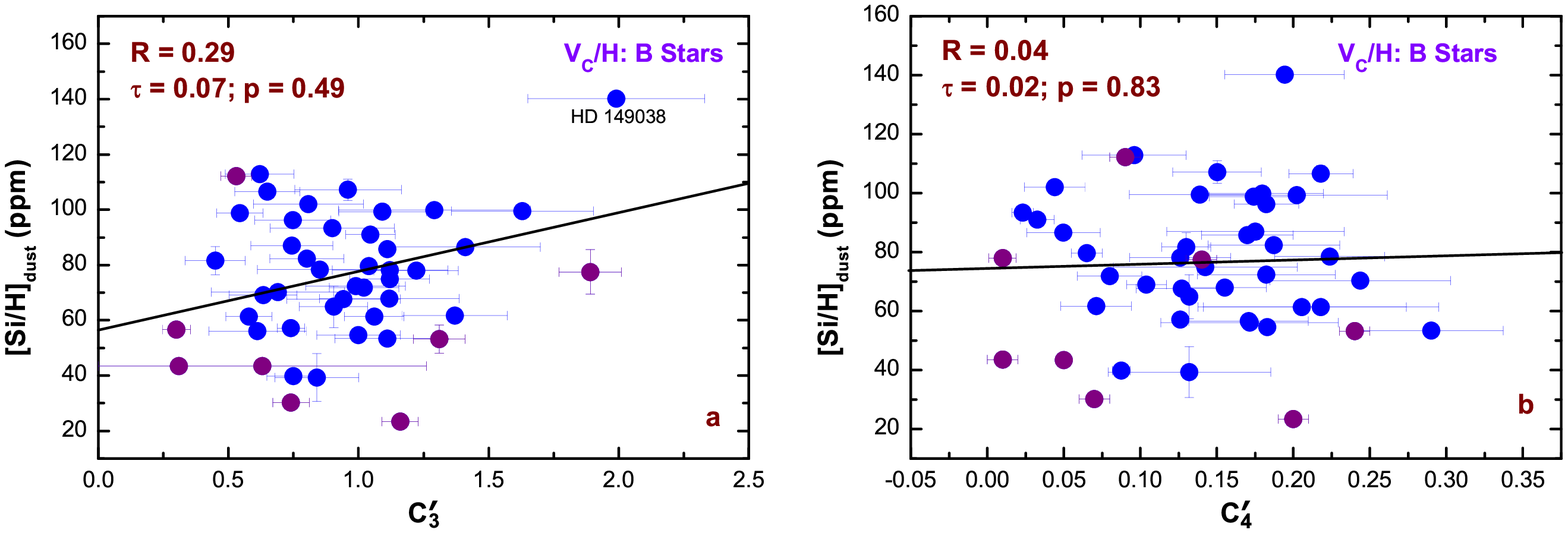}
\end{tabular}
\caption{\footnotesize
         \label{fig:c3_si2h_kk}
         Correlation diagrams between $\sidust$
         and the 2175$\Angstrom$ bump ($c_3^{\prime}$; 
	 left panel) or the far-UV rise 
	 ($c_4^{\prime}$; right panel). 
         The silicon depletion $\sidust$ is derived 
         from the Kramers-Kronig relation, 
         with the interstellar carbon reference 
         abundance $\cism$ taken to be 
         that of B stars (Przybilla et al.\ 2008)
         and the gas-phase carbon abundance $\cgas$
         derived from the hydrogen number density $\nH$
         based on the Boltzmann function
         (see Figure~\ref{fig:c2h_gas}).
         Similar to Figure~\ref{fig:c3_si2h_ism},
         the exact value of the assumed 
         interstellar C/H abundance 
         (of proto-Sun or B stars)
         does not affect
         the correlation coefficient
         but the intercept.
         }
\end{figure*}

\begin{figure*}
\centering	
\vspace{-5mm}
\includegraphics[width=1.00\textwidth]{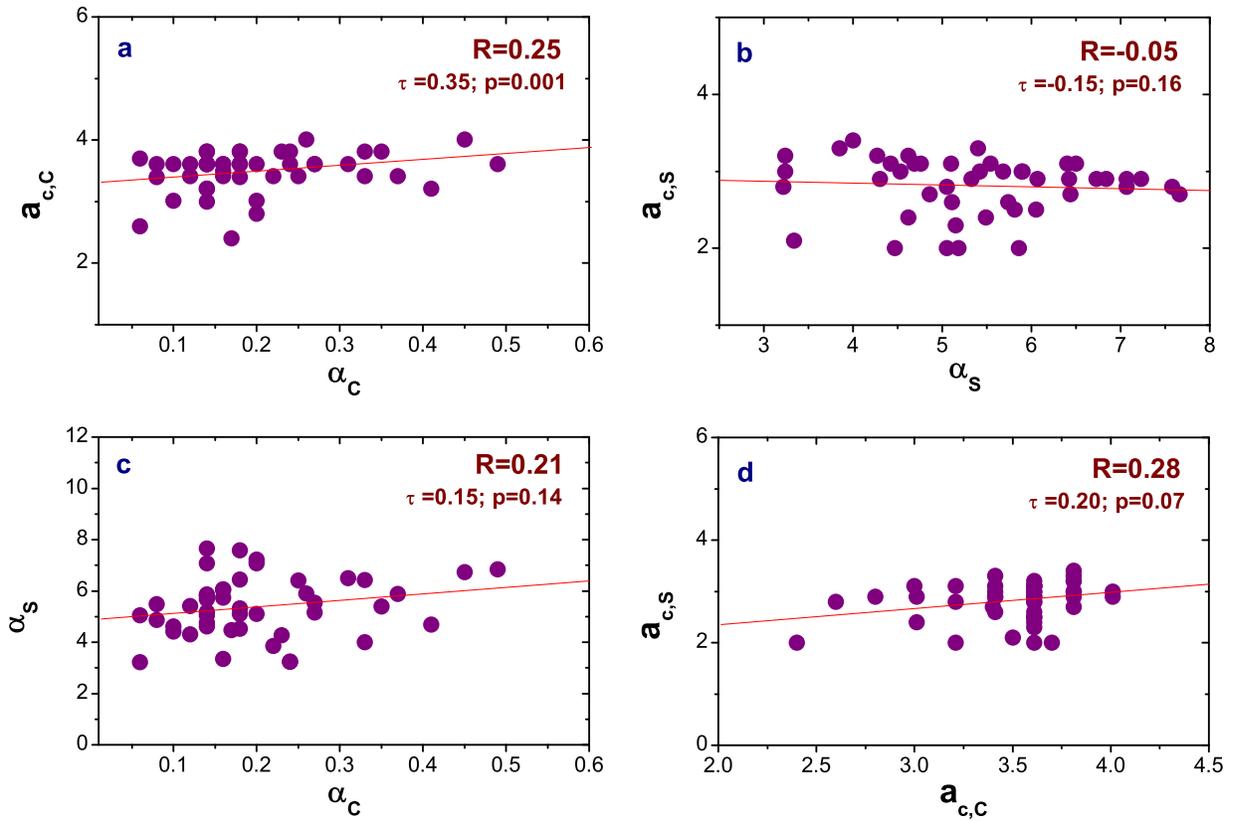}
\vspace{-2mm}
\caption{\footnotesize
        \label{fig:modpara}
        Interrelations among the model parameters
        $\acS$ -- the exponential cutoff size 
                  of silicate dust,
        $\acC$ -- the exponential cutoff size 
                  of graphite,
        $\alphaS$ -- the size distribution power 
                     index of silicate dust, and
        $\alphaC$ -- the size distribution power 
                     index of graphite.
        }
\end{figure*}

In Figures~\ref{fig:extmod1}--\ref{fig:extmod5}
we show the model fits for 39 sightlines. 
The other seven sightlines have already 
been modeled in Mishra \& Li (2015),
in a similar manner. It can be seen from these 
figures that a simple mixture of silicate and graphite
closely reproduces the observed UV/optical/near-IR 
extinction of all 46 sightlines.
The model parameters are tabulated 
in Table~\ref{tab:modpara}.
In Figure~\ref{fig:modpara} we explore
the interrelations among the dust model parameters
$\acC$, $\alphaC$, $\acS$ and $\alphaS$.
No correlations are found, 
implying that they are independent.
The other two parameters
$\Bsil$ and $\Bgra$ respectively measure 
the amounts of silicate dust and graphite dust
(see eqs.\,\ref{eq:Si2H},\,\ref{eq:C2Ha})
required to account for the observed extinction.
They are not correlated with each other or
with $\acC$, $\alphaC$, 
$\acS$ and $\alphaS$.

\begin{figure*}
\centering
\begin{tabular}{c}
\includegraphics[width=0.9\textwidth]{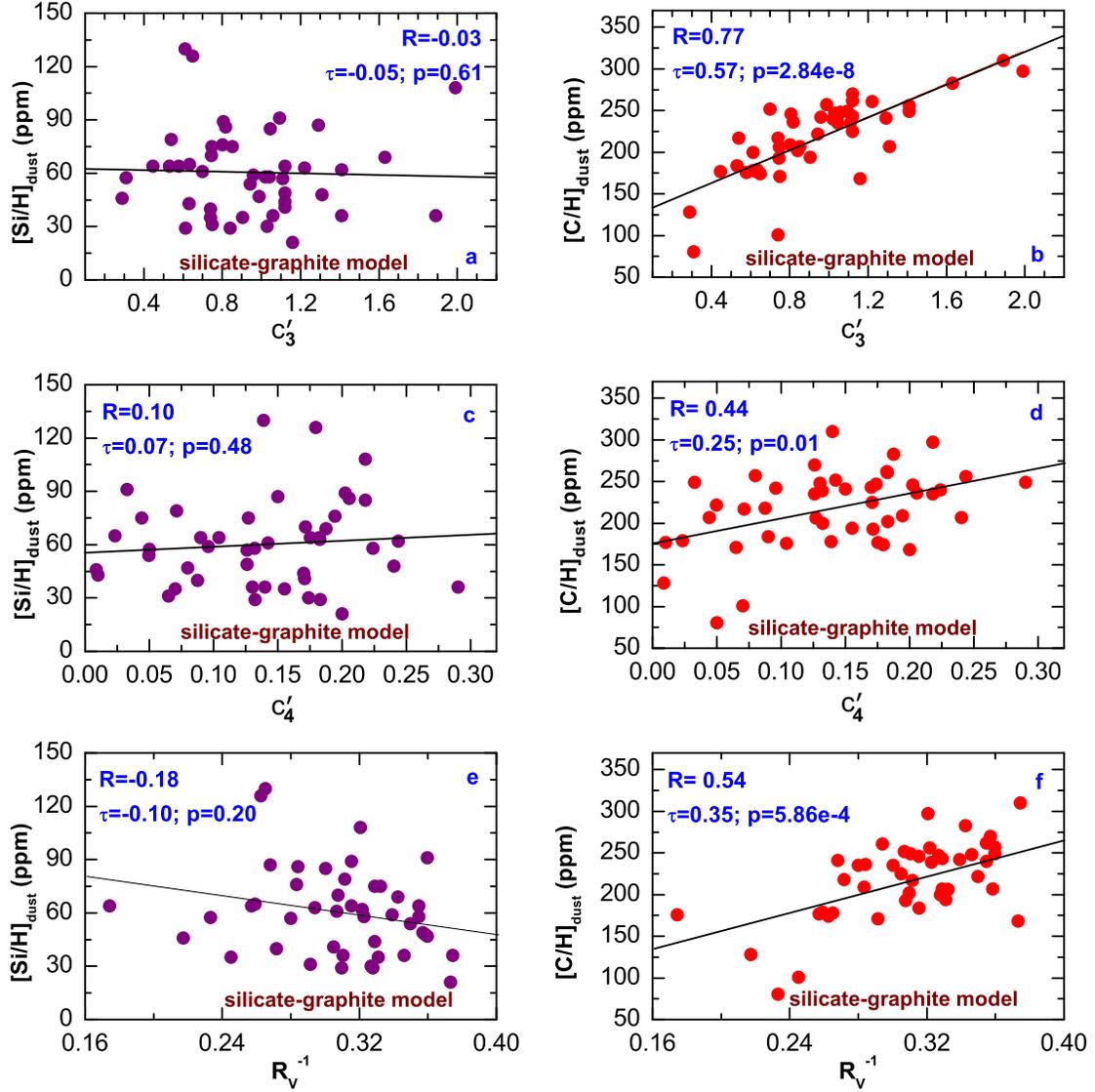}
\end{tabular}
\caption{\footnotesize
         \label{fig:c3_c2h_si2h_mod}
         Upper panel: Correlation diagrams between
         the 2175$\Angstrom$ bump ($c_3^{\prime}$) 
         and the silicon depletion $\sidust$ (a) 
         or the carbon depletion $\cdust$ (b) 
         derived from fitting the extinction curve 
         of each sightline with a mixture of
         amorphous silicate dust and graphite dust.
	 Middle panel: The correlations between
	 the far-UV nonlinear extinction 
	 rise ($c_4^{\prime}$)
         and $\sidust$ (c) or $\cdust$ (d).
	 Lower panel: The correlations between
         $R_V^{-1}$ and $\sidust$ (e) or $\cdust$ (f). 
         }
\end{figure*}

In Figure~\ref{fig:c3_c2h_si2h_mod}a
we examine the correlation between 
the strength of the 2175$\Angstrom$ 
extinction bump ($c_3^{\prime}$) 
with the silicon depletion ($\sidust$)
derived from fitting the observed extinction
(see eq.\,\ref{eq:Si2H}).
With a Pearson correlation coefficient 
of $R\approx-0.03$
and a Kendall $\tau\approx-0.05$ 
and $p\approx0.61$,
it is clear that the silicon depletion does not 
correlate with the 2175$\Angstrom$ bump. 
In contrast, the carbon depletion ($\cdust$)
derived from fitting the observed extinction
(see eq.\,\ref{eq:C2Ha}) exhibits a positive 
correlation of $R\approx0.77$, 
$\tau\approx0.57$ 
and $p\approx2.84\times10^{-8}$
with the 2175$\Angstrom$ bump
(see Figure~\ref{fig:c3_c2h_si2h_mod}b).
This is not unexpected since 
the silicate-graphite model assigns 
the 2175$\Angstrom$ bump to graphite.
As illustrated in 
Figures~\ref{fig:extmod1}--\ref{fig:extmod5},
the 2175$\Angstrom$ bump arises 
exclusively from graphite.

We have also explored the relation
between $\sidust$ and the strength of 
the nonlinear far-UV extinction rise 
($c_4^{\prime}$).
As shown in Figure~\ref{fig:c3_c2h_si2h_mod}c,
no correlation is found.
Similarly, Figure~\ref{fig:c3_c2h_si2h_mod}d
compares $\cdust$ with $c_4^{\prime}$
and also reveals no strong correlation.
However, the correlation coefficient
for $\cdust$ and $c_4^{\prime}$
($R\approx0.44$) is appreciably higher
than that for $\sidust$ and $c_4^{\prime}$
($R\approx0.10$). 
At a first glance, this is somewhat surprising 
since a visual inspection of  
Figures~\ref{fig:extmod1}--\ref{fig:extmod5}
indicates that silicates appear to
be a more important contributor 
to the far-UV extinction than graphite.
We believe that this is related to
the nature of $c_4^{\prime}$:
it arises from 
the FM mathematical separation of 
the UV extinction into 
a linear ``background'' 
at $\lambda^{-1}>3.3\mum^{-1}$ and
a nonlinear far-UV rise
at $\lambda^{-1}>5.9\mum^{-1}$.
By design, $c_4^{\prime}$ only 
measures the far-UV nonlinear rise, 
not the whole UV extinction. 
As illustrated in 
Figures~\ref{fig:extmod1}--\ref{fig:extmod5},
for many sightlines the extinction 
calculated from silicate grains 
somewhat resembles a linear ``background''. 
This explains why $\sidust$ does not
correlate with $c_4^{\prime}$. 
On the other hand, although $\cdust$
seems to show a somewhat better correlation 
with $c_4^{\prime}$, the correlation
is still very weak or at most marginal. 
This suggests that the far-UV extinction 
is more likely a combined effect of 
small silicates and small graphitic grains.

Finally, we examine how $\sidust$ and $\cdust$ 
vary with $\RV^{-1}$. 
As shown in Figures~\ref{fig:c3_c2h_si2h_mod}e,f, 
while $\sidust$ exhibits no correlation 
with $\RV^{-1}$, with a Pearson correlation 
coefficient of $R\approx0.54$,
$\cdust$ appears to moderately 
correlate with $R_V^{-1}$. 
This may merely reflect the fact that 
the 2175$\Angstrom$ bump tends 
to correlate with $R_V^{-1}$ (see Figure~7 
of Cardelli et al.\ 1989)
while the 2175$\Angstrom$ bump 
also correlates with $\cdust$
(see Figure~\ref{fig:c3_c2h_si2h_mod}b).
Cardelli et al.\ (1989) demonstrated
that not only the 2175$\Angstrom$ bump 
correlates with $R_V^{-1}$, but also 
the extinction at {\it any} other wavelengths 
within the near-IR to the far-UV 
correlates equally well with $R_V^{-1}$
(see their Figures~1,\,2).
This implies that $\cdust$ must correlate
not only with the 2175$\Angstrom$ bump,
but also and equally well with the extinction
at other wavelengths.
We stress that this is not necessarily 
inconsistent with the lack of correlation 
between $\cdust$ and $c_4^{\prime}$
(see Figure~\ref{fig:c3_c2h_si2h_mod}d)
since, as discussed above,
$c_4^{\prime}$ is not an accurate measure
of the whole UV extinction but the far-UV
nonlinear rise.
\section{Discussion}\label{sec:discussion}
In previous sections we have demonstrated that,
using both dust model-independent 
and model-dependent approaches,
the UV extinction (including the 2175$\Angstrom$ 
bump and the far-UV nonlinear rise) 
of 46 sightlines of which $\RV$ ranges 
from $\simali$2.4 to $\simali$5.8 
does not correlate with the silicon depletion.
This is in stark contrast to Haris et al.\ (2016) 
who reported a positive correlation between 
the silicon depletion and the 2175$\Angstrom$ 
bump and the far-UV rise.
Haris et al.\ (2016) derived 
such a correlation
from a smaller sample of 16 sightlines. 
They adopted
an interstellar reference abundance of
$\siism\approx33.9\ppm$ of Lodders et al.\ (2009)
and derived $\sidust$ by subtracting off 
the gas-phase abundance ($\sigas$) from $\siism$.
Our approach described in \S\ref{sec:depletion}
is similar to Haris et al.\ (2016) 
but for a larger sample of 46 sightlines.
Also, we should note that the correlations 
derived by Haris et al.\ (2016), 
with $R\approx-0.42$ and $p<0.10$
for the $\sidust$--$c_3^\prime$ relation 
and $R\approx-0.32$ and $p<0.23$
for the $\sidust$--$c_4^\prime$ relation,
were rather marginal.

\begin{figure*}
\centering
\begin{tabular}{c}
\includegraphics[width=1.0\textwidth]{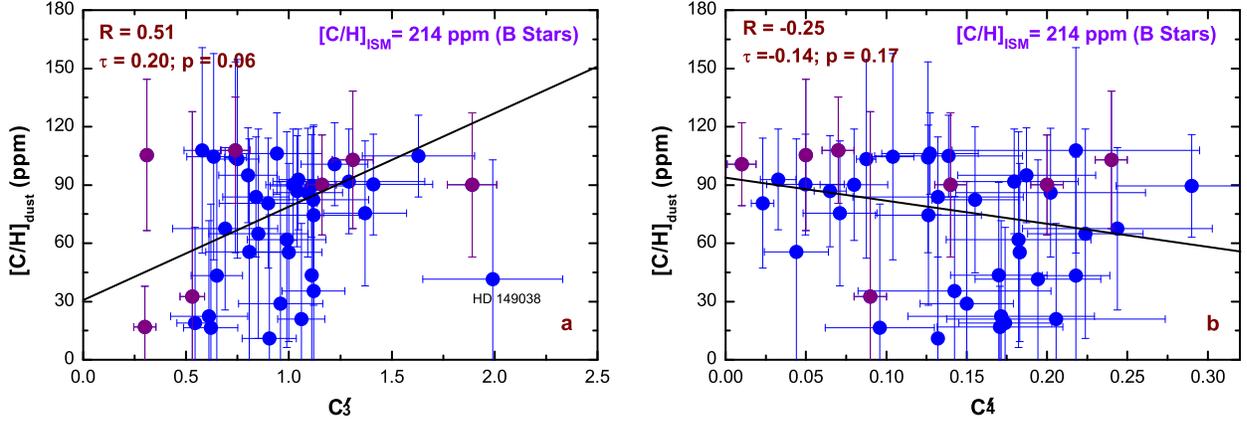}
\end{tabular}
\caption{\footnotesize
         \label{fig:c3_c2h_ism}
         Correlation diagrams 
         between the carbon depletion $\cdust$ 
         and the strength of the 2175$\Angstrom$ 
         extinction bump ($c_3^{\prime}$; left panel) 
         or the far-UV extinction rise 
	 ($c_4^{\prime}$; right panel),
         where $\cdust = \cism - \cgas$. 
         The interstellar carbon reference 
         abundance $\cism$ is assumed 
         to be that of B stars (Przybilla et al.\ 2008).
         The gas-phase carbon abundance $\cgas$
         is derived from $\nH$
         based on the Boltzmann function
         (see Figure~\ref{fig:c2h_gas}).
         Similar to Figure~\ref{fig:c3_si2h_ism},
         the exact value of the assumed 
         interstellar C/H abundance 
         does not affect
         the correlation coefficient
         but the intercept.
         }
\end{figure*}

In \S\ref{sec:extmod} we have shown that,
based on the silicate-graphite model,
the carbon depletion is correlated with
the 2175$\Angstrom$ extinction bump,
supporting the hypothesis of PAHs and/or
small graphitic grains as the carrier of
the 2175$\Angstrom$ bump
(e.g., see Stecher \& Donn 1965, 
Draine \& Malhotra 1993, Mathis 1994,
Joblin et al.\ 1992, Li \& Draine 2001a,
Cecchi-Pestellini et al.\ 2008, 
Steglich et al.\ 2010, Mulas et al.\ 2013,
Bekki et al.\ 2015). 
This is not surprising since
this model-dependent approach, 
as a priori, attributes the 2175$\Angstrom$ 
bump to graphite.\footnote{%
  The dielectric functions of 
  ``astronomical'' silicate
   and graphite of Draine \& Lee (1984) 
   adopted in \S\ref{sec:extmod}
   were designed in such a way that
   graphite causes the 2175$\Angstrom$ 
   extinction bump.
   }
To overcome this shortcoming, 
we could derive $\cdust$ 
by subtracting $\cgas$ 
(see \S\ref{sec:kk} and eq.\,\ref{eq:cgas})
from $\cism$,
with the interstellar reference abundance  
of carbon taken to be either protosolar
(Lodders 2003) or that of B stars 
(Przybilla et al.\ 2008).
This approach does not need a prior 
assignment of the 2175$\Angstrom$ bump carrier.
As illustrated in Figures~\ref{fig:c3_c2h_ism}a,
with a Pearson correlation 
coefficient of $R\approx0.51$,
$\cdust$ shows a weak
tendency of correlating with $c_3^\prime$.
This is true either the protosolar C/H abundance
or the C/H abundance of B stars is adopted
as the interstellar reference abundance.  
We note that in our previous work for 
a sample of 16 sightlines 
(see Mishra \& Li 2015),
a better correlation was derived
for $\cdust$ and $c_3^\prime$.
We speculate that the correlation 
derived here is complicated
by the fact that the gas-phase 
$\cgas$ abundance is not known
for all the 46 sighlines considered here;
instead, eq.\,\ref{eq:cgas} is used
to estimate $\cgas$ for those sightlines
for which $\cgas$ is not observationally
determined. In contrast, for the 16 lines of sight 
considered in Mishra \& Li (2015), 
the gas-phase $\cgas$ abundance
is known for every line of sight.

Alternatively, we can also derive 
the carbon depletion $\cdust$ 
required to account for the observed extinction
from the KK relation of Purcell (1969).
Similar to the approach described in \S\ref{sec:kk},
by assuming an interstellar silicon reference abundance 
of $\siism$, for a given sightline of known $\sigas$ 
we first derive the total volume (per H nucleon) 
of silicate dust from
\begin{equation}\label{eq:Vsil2}
 \frac{\Vsil}{\rm H} = 
\left\{\siism-\sigas\right\}\times\mu_{\rm sil}\mH/\rho_{\rm sil} ~~,
\end{equation}
where $\mu_{\rm sil}=172$ if we assume 
a stoichiometric composition of MgFeSiO$_4$. 
By applying the KK relation of Purcell (1969), 
we obtain the wavelength-integrated extinction of 
silicate origin from the silicate dust volume through 
\begin{equation}\label{eq:kkc}
\int_{0}^{\infty} \left(\frac{A_\lambda}{\NH}\right)_{\rm sil} d\lambda 
= 1.086\times 3 \pi^2 \Fsil \frac{\Vsil}{\rm H} ~~.
\end{equation}
Again, if we assume amorphous silicate and 
carbon dust as two major dust populations 
in the ISM, we derive the extinction contributed 
by carbon dust from 
\begin{equation}\label{eq:AC}
\int_{0}^{\infty} \left(\frac{A_\lambda}{\NH}\right)_{\rm C} d\lambda 
= \int_{0}^{\infty} \left(\frac{A_\lambda}{\NH}\right)_{\rm obs} d\lambda 
- \int_{0}^{\infty} \left(\frac{A_\lambda}{\NH}\right)_{\rm sil} d\lambda ~~.
\end{equation}
We then apply the KK relation again to deduce
the total volume of carbon dust required to
account for the carbon-originated extinction
$\int_{0}^{\infty} \left(A_\lambda/\NH\right)_{\rm C} d\lambda$: 
\begin{equation}\label{eq:VC}
\frac{\VC}{\rm H}  = 
\frac
{\int_{0}^{\infty} \left(A_\lambda/\NH\right)_{\rm C} d\lambda} 
{1.086\times 3 \pi^2 \FC} ~~.
\end{equation}
Finally, from the carbon dust volume 
we derive the carbon depletion to be 
\beq\label{eq:C2Hb}
\cdust = \frac
{\left(\VC/\rmH\right)\rho_{\rm C}}
{12\mH} ~~.
\eeq
This approach does not require the knowledge
of the unknown interstellar carbon reference 
abundance $\cism$ and the gas-phase carbon
abundance $\cgas$. The latter is also unknown 
for some sightlines and in previous sections, it
was estimated from $\nH$ through eq.\,\ref{eq:cgas}.

\begin{figure*}
\centering
\begin{tabular}{c}
\includegraphics[width=1.0\textwidth]{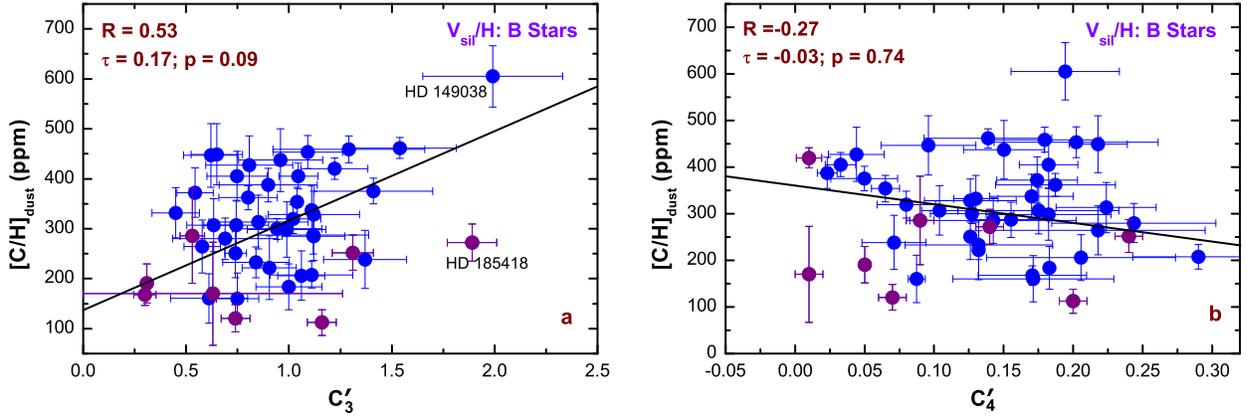}
\end{tabular}
\caption{\footnotesize
         \label{fig:c3_c2h_kk}
         Correlation diagrams between $\cdust$
         and the 2175$\Angstrom$ bump ($c_3^{\prime}$; 
	 left panel) or the far-UV extinction 
	 rise ($c_4^{\prime}$; right panel).
         The carbon depletion $\cdust$ is derived 
         from the Kramers-Kronig relation, 
         with the interstellar Fe, Mg, and Si reference 
         abundances taken to be 
         that of B stars (Przybilla et al.\ 2008)
         and the gas-phase silicon abundance $\sigas$
         taken from Haris et al.\ (2016).
         Similar to Figure~\ref{fig:c3_si2h_ism},
         the exact values of the assumed 
         interstellar Fe/H, Mg/H and Si/H abundances 
         do not affect
         the correlation coefficient
         but the intercept.
         }
\end{figure*}

For each sightline, assuming the interstellar silicon
reference abundance to be either protosolar
($\siism=\sisun\approx40.7\pm1.9\ppm$; Lodders 2003) 
or like B stars ($\siism=\sistar\approx31.6\pm1.5\ppm$;
Przybilla et al.\ 2008) and taking the observed extinction 
$\left(A_\lambda/\NH\right)_{\rm obs}$ constructed
in \S\ref{sec:extmod} and the observationally
determined gas-phase silicon abundance $\sigas$,
we derive the carbon depletion $\cdust$ from
eqs.\,\ref{eq:Vsil2}--\ref{eq:C2Hb} and then compare 
with the 2175$\Angstrom$ extinction bump
(measured by $c_3^{\prime}$) .
As shown in Figures~\ref{fig:c3_c2h_kk}a,
with a Pearson correlation coefficient of 
$R\approx0.53$, $\cdust$ tends to weakly 
correlate with $c_3^{\prime}$.
Again, this is true 
either the protosolar Si/H abundance
or the Si/H abundance of B stars is adopted
as the interstellar reference abundance.  
This supports the hypothesis of some sorts of 
carbonaceous grains (e.g., graphite or PAHs)
as the possible carriers of the 2175$\Angstrom$ bump,
while the lack of correlation between the silicon depletion
and the 2175$\Angstrom$ bump (see \S\ref{sec:depletion},
\S\ref{sec:kk}, and \S\ref{sec:extmod}) argues against
small silicates or (Mg, Si) oxides as its carrier 
(Duley 1985, Steel \& Duley 1987,
Parvathi et al.\ 2012, Haris et al.\ 2016).  

We also show in Figure~\ref{fig:c3_c2h_ism}b 
and Figure~\ref{fig:c3_c2h_kk}b that
$\cdust$ does not correlate with 
the far-UV extinction rise as measured 
by $c_4^{\prime}$.
This, together with the lack of correlation
of $\sidust$ with $c_4^{\prime}$ 
(see \S\ref{sec:depletion}, \S\ref{sec:kk},
and \S\ref{sec:extmod}), suggests that
neither small silicate grains 
nor small carbon grains alone
account for the far-UV extinction rise,
instead, it must be their combined effects.
Indeed, as illustrated in Figure~3 of
Xiang et al.\ (2017), the silicate-graphite-PAH 
model of WD01 requires both silicate grains of
sizes $a\simlt 250\Angstrom$ and 
graphitics grains or PAHs 
of $a\simlt 250\Angstrom$ 
to appreciably contribute to the far-UV extinction.
Moreover, the nondetection of correlation 
between the 2175$\Angstrom$ bump
and the far-UV rise (see Greenberg \& Chlewicki 1983,
Rouleau et al.\ 1997, Xiang et al.\ 2017)
also implies that the bump carriers are not 
a dominant contributor of the far-UV extinction.
Finally, we note that, while the FM parametrization 
provides an excellent mathematical description
of the UV extinction at $\lambda^{-1}>3.3\mum^{-1}$,
the distinction between the linear rise 
(measured by $c_1^{\prime}$ and $c_2^{\prime}$)
and the nonlinear far-UV rise (measured by $c_4^{\prime}$)
probably has little physical significance 
since there is no substance known that shows 
the corresponding extinction of any of them.
We suggest that the decomposition scheme
originally proposed by Greenberg (1973) and
very recently revisited by Xiang et al.\ (2017)
may be a better characterization of the far-UV extinction.
According to this decomposition scheme, 
any observed interstellar extinction curve
can be decomposed into three parts:
(i) a near-IR/visible component 
which flattens off in the UV and far-UV,
(ii) a bump at 2175$\Angstrom$, and 
(iii) a far-UV component.
It would be interesting to explore the relations
of the carbon and silicon depletions 
with these decomposed extinction components.

To summarize, we have demonstrated 
the refutation of Haris et al.\ (2016)
by expanding our earlier study (Mishra \& Li 2015) 
to provide a better understanding 
of the extinction curve and show that 
the 2175$\Angstrom$ extinction does not 
correlate with the silicon depletion. 
We should stress that, although this study 
supports the hypothesis of graphite or PAHs 
as the possible carriers of 
the 2175$\Angstrom$ bump, it has not yet 
necessarily approved this hypothesis. 
Neither the carrier of the bump 
nor those of the far-UV rise are yet assigned.
The attribution of the far-UV extinction to 
a mixture of small carbon grains 
and small silicates also remains hypothetical.
The 2175$\Angstrom$ bump could well be
caused by a separate population of grains 
other than graphite or PAHs, 
e.g., carbon buckyonions (Chhowalla et al.\ 2003, 
Iglesias-Groth et al.\ 2003,
Ruiz et al.\ 2005, Li et al.\ 2008).
The far-UV extinction could also be 
partly contributed by iron nanoparticles
(e.g., see Hensley \& Draine 2017b).
Also, the division of the UV extinction curve 
into a three-component scheme is 
by no means physical. 

\begin{figure*}
\centering	
\vspace{-5mm}
\includegraphics[width=1.00\textwidth]{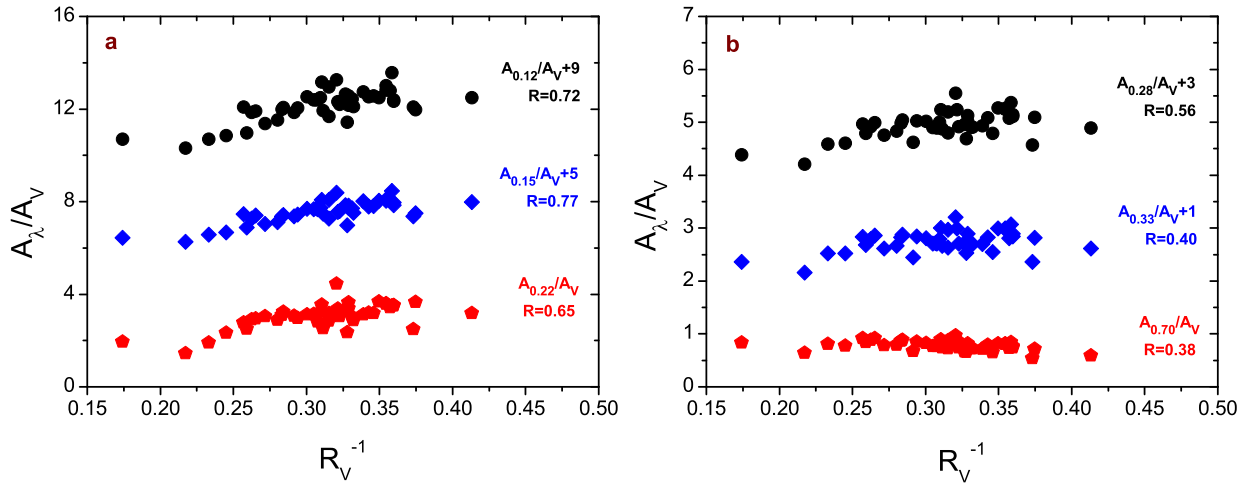}
\vspace{-2mm}
\caption{\footnotesize
         \label{fig:Alambda_RV}
         The extinction ratio $A_\lambda/A_V$ 
         plotted against $\RV^{-1}$ 
         at selected wavelengths.
         The subscripts refer to the wavelength
         (e.g., $A_{0.12}$ refers to the extinction
         at $\lambda=0.12\mum$).
         The data for $\lambda=0.12, 0.15, 0.28, 0.33\mum$
         have been shifted vertically by the amount
         indicated in order to separate them.
         }
\end{figure*}

Finally, we show in Figure~\ref{fig:Alambda_RV}
the variations of $A_\lambda/A_V$ with $R_V^{-1}$
for $\lambda=0.12\mum$, 0.15$\mum$, 0.22$\mum$, 
0.28$\mum$, 0.33$\mum$, and 0.70$\mum$.
One can see that at $\lambda=0.12\mum$, 
0.15$\mum$, 0.22$\mum$, and 0.28$\mum$, 
there is a good linear relationship 
between $A_\lambda/A_V$ and $R_V^{-1}$,
confirming the earlier findings of 
Cardelli et al.\ (1989).
At $\lambda=0.33\mum$ and 0.70$\mum$,
$A_\lambda/A_V$ appears to correlate
with $R_V^{-1}$ much more weakly
(if at all). This could be related 
to the use of $A_V$ for normalizing 
the extinction. As illustrated in Figure~2
of Cardelli et al.\ (1989), 
if normalized to the $I$-band extinction 
at $\lambda\approx0.90\mum$,
the extinction, expressed as $A_\lambda/A_I$,
shows a clear relation with $R_V^{-1}$ for all
wavelengths at $\lambda<0.90\mum$.\footnote{%
  The extinction law at $\lambda\simgt0.90\mum$
  appears to vary very little with environments
  (e.g., see Martin \& Whittet 1990, 
   Wang \& Jiang 2014, Wang et al.\ 2013).
  }

The correlation between $R_V^{-1}$ 
and the extinction at {\it any} wavelength 
from the optical to the far-UV
suggests the existence of 
a common process that simultaneously 
modifies all parts of the extinction curve
(see \S4 in Cardelli et al.\ 1989).
As the extinction at 
a particular wavelength $\lambda$
is dominated by grains of 
a particular size $a\sim\lambda/2\pi$
(see Li 2009), the fact that $A_\lambda/A_V$
correlates with $R_V^{-1}$ at all wavelengths
implies that the process which produces changes
in extinction must operate effectively and rather
continuously over most of the range of 
grain sizes (except the largest sizes 
since the near- and mid-IR extinction 
seems to show little dependence with 
environments; see Wang et al.\ 2014,
Xue et al.\ 2016).
To examine this, 
we explore
how the mean grain sizes derived 
in \S\ref{sec:extmod}
from fitting the observed extinction curves
vary with $R_V^{-1}$.

Let $\ameansil$ and $\ameangra$
respectively be the mean sizes 
of the silicate and graphite grains.
For a given weighting factor $\omega(a)$,
we derive $\ameansil$ and $\ameangra$ 
from the size distribution parameters
determined in \S\ref{sec:extmod} 
as follows:
\begin{equation}\label{eq:ameansilgra}
\ameansil = \frac{\int da\,\omega(a)\,a^{1-\alphaS}
            \exp\left(-a/\acS\right)}
            {\int da\,\omega(a)\,a^{-\alphaS}
            \exp\left(-a/\acS\right)} ~,~~
\ameangra = \frac{\int da\,\omega(a)\,a^{1-\alphaC}
            \exp\left(-a/\acC\right)}
            {\int da\,\omega(a)\,a^{-\alphaC}
            \exp\left(-a/\acC\right)} ~.
\end{equation}
We shall consider three kinds of weighting factors:
$\omega(a)=a^2$ 
(i.e., weighted by grain surface area),
$\omega(a)=a^3$ 
(i.e., weighted by grain mass or volume), and
$\omega(a)=\Cext(a,\lambda)$ 
(i.e., weighted by extinction cross section 
at wavelength $\lambda$).
We define the overall mean grain size 
as the average of $\ameansil$ and $\ameangra$,
weighted by the mass fraction 
of each dust component:
\begin{equation}\label{eq:amean}
\amean = \ameansil  
         \frac{\musil\sidust}
         {\muc\cdust + \musil\sidust} 
       + \ameangra  
         \frac{\muc\cdust}
         {\muc\cdust + \musil\sidust} ~~, 
\end{equation}
where the silicon ($\sidust$) 
and carbon depletions ($\cdust$) 
are determined in \S\ref{sec:extmod} 
from modeling the observed extinction curves
(see eqs.\,\ref{eq:Si2H},\,\ref{eq:C2Ha}
and Table~\ref{tab:modpara}).

\begin{figure*}
\centering	
\vspace{-5mm}
\includegraphics[width=1.00\textwidth]{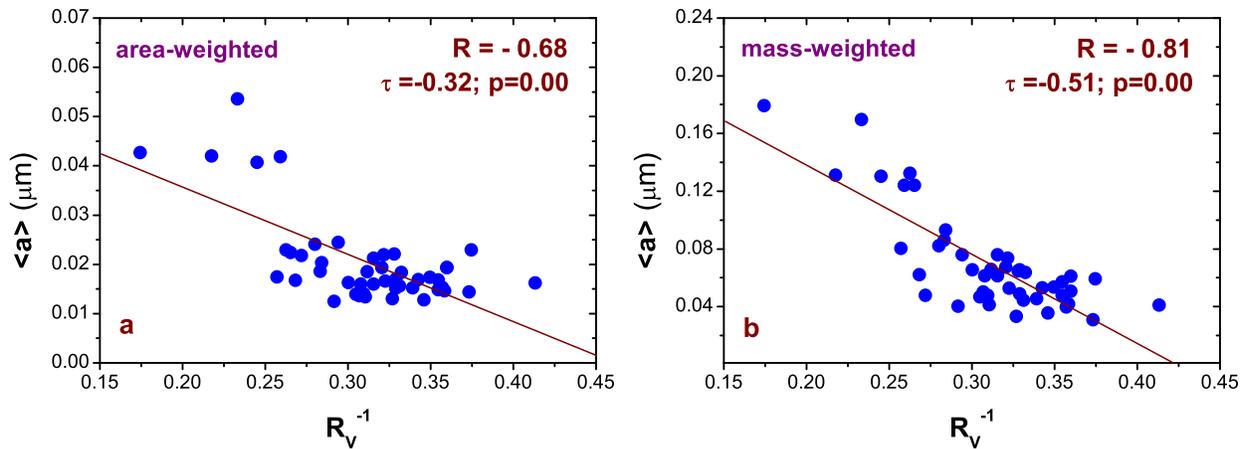}
\vspace{-2mm}
\caption{\footnotesize
        \label{fig:<a>_area_mass}
         Area-weighted (a) and
         mass-weighted (b) 
         mean grain sizes vs. $\RV^{-1}$.
         }
\end{figure*}

\begin{figure*}
\centering	
\vspace{-5mm}
\includegraphics[width=1.00\textwidth]{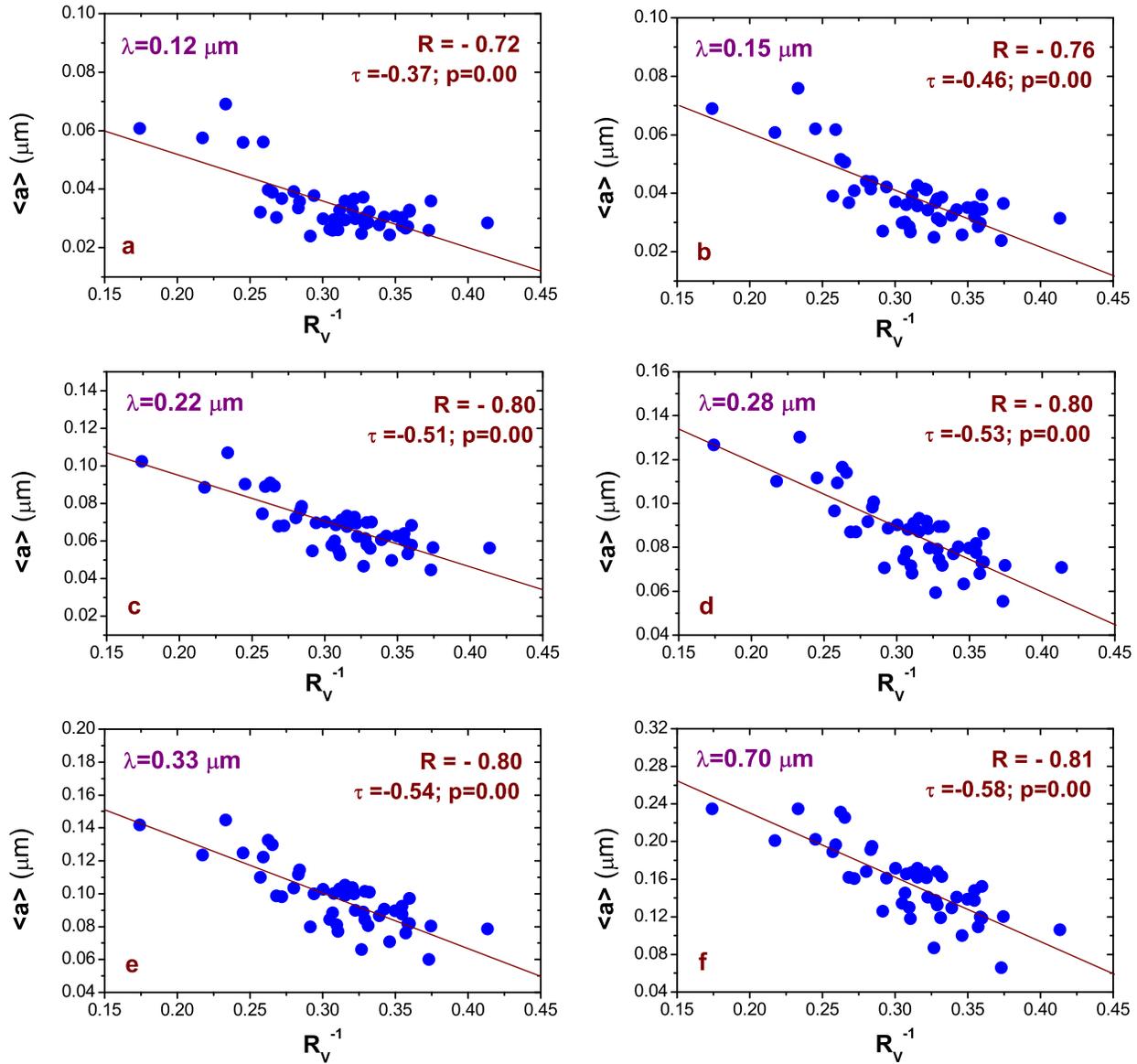}
\vspace{-2mm}
\caption{\footnotesize
         \label{fig:<a>_ext}
         Extinction-weighted 
         mean grain sizes
         vs. $\RV^{-1}$.
         The weighting extinction cross sections
         $C_{\rm ext}(a,\lambda)$
         are taken to be that at 
         $\lambda=0.12\mum$ (a), 
         0.15$\mum$ (b), 0.22$\mum$ (c), 
         0.28$\mum$ (d), 0.33$\mum$ (e),
         and 0.70$\mum$ (f).
         }
\end{figure*}

In Figure~\ref{fig:<a>_area_mass}
we show the area- and mass-weighted 
mean grain sizes ($\amean$)
as a function of $R_V^{-1}$.
One can see that the mean sizes 
clearly anti-correlate with $R_V^{-1}$ 
(i.e., in denser regions of 
larger $R_V$ values, on an average,
the grains are larger). 
As illustrated in Figure~\ref{fig:<a>_ext},
the anti-correlation 
between $\amean$ and $R_V^{-1}$
is also seen for 
the $\Cext(a,\lambda)$-weighted 
mean grain sizes, with $\Cext(a,\lambda)$
calculated at various wavelengths
from the optical to the far-UV.
This demonstrates that whatever processes
modify the dust size distribution in one regime 
and the extinction at one wavelength
must act in a rather systematic fashion 
over the entire size distribution and
the extinction over the entire wavelength 
range from the optical to the far-UV.
However, it is not clear what processes
play a dominant role in regulating
the extinction curve and the dust sizes
(see Cardelli et al.\ 1989).

\section{Conclusion}\label{sec:conclusion}
We have studied the extinction and dust depletion
in 46 Galactic sightlines of which $\RV$ ranges 
from $\simali$2.4 to $\simali$5.8 
in order to probe the role of silicon 
in the UV extinction, particularly
the 2175$\Angstrom$ extinction bump.
These sightlines, with their UV/optical/near-IR
extinction, gas-phase silicon abundances, and 
hydrogen column densities observationally determined,
allow us to quantitatively explore the relations 
between the silicon depletion and the UV extinction.
Our principal results are as follows:
\begin{enumerate}
\item By deriving the silicon depletion $\sidust$ 
          from subtracting the observed gas-phase 
          silicon abundance $\sigas$
          from the assumed interstellar silicon reference 
          abundance $\siism$, we find that $\sidust$ is 
          not correlated either with the 2175$\Angstrom$ 
          extinction bump or with the far-UV extinction rise. 
          This approach is independent of any dust models.
\item By deriving $\sidust$ from the Kramers-Kronig relation 
          which relates the wavelength-integrated extinction
          to the total dust volume, we also find no correlation
          between $\sidust$ and the 2175$\Angstrom$ bump
          or the far-UV extinction rise. This approach is also
          model independent in the sense that it does not require 
          the knowledge of the exact optical properties, composition
          and size distribution of the dust. 
\item We have also derived $\sidust$ as well as 
          the carbon depletion $\cdust$ from fitting 
          the observed UV/optical/near-IR extinction 
          with a mixture of silicate dust and carbon dust.
          While no correlation is found 
          between $\sidust$ and the 2175$\Angstrom$ bump 
          or between $\sidust$ and the far-UV rise,
          $\cdust$ does show a positive correlation
          with the 2175$\Angstrom$ bump.
\item We have also derived $\cdust$ either from
          subtracting the gas-phase abundance $\cgas$
          from the assumed interstellar reference 
          abundance $\cism$
          or from the Kramers-Kronig relation. 
          In both cases we find that $\cdust$ 
          tends to correlate with 
          the 2175$\Angstrom$ bump
          (but not with the far-UV rise).
\item We conclude that some sorts of 
          carbonaceous grains (e.g., graphite or PAHs)
          are the most plausible carrier of 
          the 2175$\Angstrom$ bump,
          and silicates or (Mg, Si) oxides  
          are unlikely responsible for the bump.
          Neither small silicate grains 
          nor small carbon grains alone
          account for the the far-UV extinction rise,
          instead, it must be their combined effects.
\item We have shown that the extinction 
           at {\it any} wavelength
           from the optical to the far-UV 
           correlates with $R_V^{-1}$,
           consistent with the earlier findings of 
           Cardelli et al.\ (1989).
           We have also found that the area-, mass-,
           or extinction-weighted mean grain size
           averaged over silicate dust and graphite dust
           anti-correlates with $R_V^{-1}$.
           This demonstrates that in the ISM
           the processes which modify the extinction
           at one wavelength and the grain size 
           in one regime must also modify the extinction
           over the entire wavelength range 
           and the grain size over the entire size range,
           although the exact processes remain unknown.
\end{enumerate}

\acknowledgments
We thank B.T.~Draine, B.W.~Jiang, H.~Kimura,
A.N.~Witt, G.~Zhao and the anonymous referee 
for helpful discussions and comments. 
We are supported in part by NSF AST-1109039, 
NNX13AE63G, and NSFC\,11173019, 11273022
and the University of Missouri Research Board.


\clearpage

\clearpage

\clearpage

\clearpage
\thispagestyle{empty}
\setlength{\voffset}{25mm}
\begin{deluxetable}{lccccccccccccccccccccccr}
\rotate 
\tablecolumns{17}
\tabcolsep=0.06cm
\tabletypesize{\scriptsize}
\tablewidth{0truein}
\center
\tablecaption{\footnotesize
              \label{tab:kk}
              Extinction Parameters 
              and the Silicon and Carbon Depletions
              Required to Account for
              the Observed Extinction
              }
\tablehead{
\colhead{Star}&
\colhead{$N_{\rm H}^{a}$}&
\colhead{$\log_{10}\left(\nH\right)^{b}$}&
\colhead{$R_{V}$}&
\colhead{$A_{V}$}&
\colhead{$c_{1}^{\prime}$}&
\colhead{$c_{2}^{\prime}$}&
\colhead{$c_{3}^{\prime}$}&
\colhead{$c_{4}^{\prime}$}&
\colhead{$x_{0}$}&
\colhead{$\gamma$}&
\colhead{$A_{\rm int}^{c}$}&
\colhead{}&
\colhead{Proto-Sun$^{d}$}&
\colhead{}&
\colhead{}&
\colhead{}&
\colhead{}&
\colhead{}&
\colhead{B Stars$^{e}$}&
\colhead{}&
\colhead{}
\\
\cline{13-16}  
\cline{18-21}
\\
\colhead{HD}&
\colhead{}&
\colhead{$({\rm cm}^{-3})$}&
\colhead{}&
\colhead{(mag)}&
\colhead{}&
\colhead{}&
\colhead{}&
\colhead{}&
\colhead{($\um^{-1}$)}&
\colhead{($\um^{-1}$)}&
\colhead{}&
\colhead{$\Vsil/\rmH^{f}$}&
\colhead{$\stohd^{g}$}&
\colhead{$\VC/\rmH^{h}$}&
\colhead{$\cdust^{i}$}&
\colhead{}&
\colhead{$\Vsil/\rmH^{f}$}&
\colhead{$\stohd^{g}$}&
\colhead{$\VC/\rmH^{h}$}&
\colhead{$\cdust^{i}$}
}
\startdata

001383$^{1}$ & 3.16$^{+0.64}_{-0.53}$&-0.45 &3.04$\pm$0.15$^{2}$ & 1.43$\pm$0.14 & 1.13$\pm$0.31 & 0.21$\pm$0.02 &	1.13$\pm$0.15	& 0.17$\pm$0.03 & 4.60$\pm$0.01 & 0.91$\pm$0.03 & 1.50 & 4.85E-27 & 59$\pm$2	&	2.12E-27 & 238$\pm$54 && 6.11E-27 & 75$\pm$2 & 2.54E-27	& 285$\pm$49 \\
002905$^{1}$ & 2.00 & -0.22& 3.26$\pm$0.20$^{5}$ & 1.08$\pm$0.12 & 1.11$\pm$0.80&0.26$\pm$0.09& 0.69$\pm$0.26& 0.14$\pm$0.06& 4.60$\pm$0.02&0.82$\pm$0.03&1.53& 4.47E-27	&55$\pm$2& 2.07E-27& 233$\pm$48	&&	5.73E-27&70$\pm$2	&2.49E-27& 280$\pm$42	\\
023180$^{1}$ & 1.62 &0.065&3.11$\pm$0.39$^{1}$ &0.93$\pm$0.17&1.49$\pm$0.53&0.10$\pm$0.02&1.41$\pm$0.29&0.24$\pm$0.06& 4.57$\pm$0.02&	1.11$\pm$0.04	&1.91	&5.80E-27&71$\pm$2&2.92E-27&329$\pm$35	&&	7.06E-27&87$\pm$2	&3.34E-27&375$\pm$26	\\
024912$^{1}$ & 1.98$^{+0.54}_{-0.54}$ &0.75&2.86$\pm$0.51$^{1}$ &1.00$\pm$0.21&1.19$\pm$0.73	&0.27$\pm$0.05	&0.94$\pm$0.22	&0.05$\pm$0.02 &4.54$\pm$0.02	&0.85$\pm$0.03	&1.62	&4.26E-27&52$\pm$2&	2.25E-27	&253$\pm$31&&	5.52E-27	& 68$\pm$2&	2.66E-27&	300$\pm$21	\\
030614$^{1}$ & 1.23 &-0.64& 3.01$\pm$0.33$^{1}$	&0.87$\pm$0.15	&1.10$\pm$0.29	&0.20$\pm$0.03	&0.75$\pm$0.15	&0.13$\pm$0.03	&	4.57$\pm$0.01	&	0.90$\pm$0.03	&	1.75	&	6.60E-27	&	81$\pm$3	&	3.19E-27	&	358$\pm$55	&&	7.86E-27	&	96$\pm$2	&	3.60E-27	&	405$\pm$50	\\
034078$^{1}$ & 3.10 & 0.35&	3.40$\pm$0.22$^{1}$	&	1.80$\pm$0.18	&	1.37$\pm$0.20	&	0.14$\pm$0.02	&	1.22$\pm$0.16	&	0.18$\pm$0.02	&	4.60$\pm$0.00	&	1.11$\pm$0.03	&	1.79	&	5.10E-27	&	63$\pm$2	&	3.32E-27	&	373$\pm$32	&&	6.36E-27	&	78$\pm$2	&	3.73E-27	&	420$\pm$21 \\
037021$^{1}$ & 4.79$^{+1.52}_{-1.16}$ &0.61&	4.60$\pm$0.12$^{6}$	&	2.48$\pm$0.06	&	1.08$\pm$1.06	&	0.03$\pm$0.01	&	0.30$\pm$0.05	&	0.01$\pm$0.01	&	4.58$\pm$0.05	&	1.08$\pm$0.04	&	1.10	&	3.36E-27	&	41$\pm$2	&	1.07E-27	&	121$\pm$31	&&	4.62E-27	&	57$\pm$2	&	1.49E-27	&	168$\pm$21	\\
041117$^{1}$ & 3.47$^{+1.43}_{-0.90}$ &0.05&	3.28$\pm$0.14$^{5}$	&	1.48$\pm$0.06	&	1.04$\pm$0.36	&	0.23$\pm$0.03	&	1.30$\pm$0.20	&	0.23$\pm$0.04	&	4.61$\pm$0.01	&	0.99$\pm$0.03	&	1.30	&	3.10E-27	&	38$\pm$2	&	1.43E-27	&	161$\pm$35	&&	4.36E-27	&	53$\pm$2	&	1.85E-27	&	208$\pm$27	\\
042087$^{1}$ & 3.09$^{+0.71}_{-0.58}$ &-0.14&	3.22$^{7}$		&	1.19$\pm$0.13	&	1.08$\pm$0.69	&	0.26$\pm$0.04	&	1.38$\pm$0.20	&	0.29$\pm$0.05	&	4.61$\pm$0.00	&	1.07$\pm$0.02	&	1.40	&	3.77E-27	&	46$\pm$2	&	1.70E-27	&	192$\pm$63	&&	5.03E-27	&	62$\pm$2	&	2.12E-27	&	238$\pm$58	\\
045314$^{1}$ & 1.91 &-0.53&	3.21$\pm$0.14$^{5}$	&	1.38$\pm$0.14	&	1.09$\pm$0.22	&	0.20$\pm$0.02	&	0.54$\pm$0.09	&	0.07$\pm$0.02	&	4.59$\pm$0.02	&	0.90$\pm$0.03	&	1.88	&	6.80E-27	&	83$\pm$2	&	2.89E-27	&	325$\pm$55	&&	8.06E-27	&	99$\pm$2	&	3.31E-27	&	372$\pm$49	\\
046056$^{1}$ & 3.39$^{+0.98}_{-0.76}$ &-0.32&	3.06$\pm$0.11$^{5}$	&	1.50$\pm$0.13	&0.79$\pm$0.34	&	0.29$\pm$0.03	&	1.04$\pm$0.16	&	0.18$\pm$0.03	&	4.58$\pm$0.01	&0.91$\pm$0.03	&	1.20	&	3.20E-27	&	39$\pm$2	&	1.22E-27	&	137$\pm$51	&&	4.46E-27	&	55$\pm$2	&	1.64E-27	&	184$\pm$46	\\
046202$^{3}$ & 4.79$^{+1.67}_{-1.16}$ &-0.04&	3.23$\pm$0.14$^{5}$	&	1.55$\pm$0.15	&	1.03    	&	0.23		       &	0.84		&	0.18		&	4.59	                	&	0.86	        	&	1.02	&	1.94E-27	&	24$\pm$9	&	1.65E-27	&	186$\pm$39	&&	3.20E-27	&	39$\pm$9	&	2.07E-27	&	233$\pm$31	\\
069106$^{1}$ & 1.32$^{+0.19}_{-0.14}$ &-0.52&	3.05$\pm$0.44$^{1}$	&	0.61$\pm$0.15	&	1.27$\pm$0.65	&	0.10$\pm$0.03	&	0.61$\pm$0.19	&	0.13$\pm$0.05	&	4.59$\pm$0.02	&	0.96$\pm$0.03	&	1.11	&	3.32E-27	&	41$\pm$2	&	1.01E-27	&	114$\pm$55	&&	4.58E-27	&	56$\pm$2	&	1.43E-27	&	161$\pm$49	\\
093205$^{1}$ & 2.24 &-0.66&	3.25$\pm$0.24$^{1}$	&	1.24$\pm$0.16	&	0.93$\pm$0.51	&	0.25$\pm$0.03	&	0.74$\pm$0.16	&	0.17$\pm$0.06	&	4.61$\pm$0.04	&	0.96$\pm$0.03	&	1.56	&	5.83E-27	&	71$\pm$3	&	2.31E-27	&	260$\pm$55	&&	7.09E-27	&	87$\pm$3	&	2.73E-27	&	307$\pm$50	\\
093843$^{1}$ &	2.24$^{+0.45}_{-0.38}$	&-0.68&	3.89$\pm$0.41$^{1}$	&	1.05$\pm$0.22	&	1.37$\pm$0.50	&	0.15$\pm$0.03	&	0.45$\pm$0.11	&	0.18$\pm$0.06	& 4.57$\pm$0.03	&	0.78$\pm$0.03	&	1.44	&	5.40E-27	&	66$\pm$5	&	2.54E-27	&	285$\pm$55	&&	6.66E-27	&	82$\pm$5	&	2.95E-27	&	332$\pm$50	\\
101190$^{2}$ & 1.95 &-0.52&	2.89$\pm$0.39$^{2}$	&	0.90$\pm$0.13	&	0.44$\pm$0.09	&	0.34$\pm$0.02	&	1.06$\pm$0.11	&	0.13$\pm$0.02	&4.59$\pm$0.05	&	0.93$\pm$0.02	&	1.20	&	3.74E-27	&	46$\pm$2	&	1.42E-27	&	159$\pm$55	&&	5.00E-27	&	61$\pm$2	&	1.83E-27	& 206$\pm$49	\\
112244$^{1}$ & 1.48 &-0.08&	3.52$\pm$0.16$^{5}$	&	0.95$\pm$0.32	&	1.50$\pm$0.49	&	0.11$\pm$0.03	&	0.89$\pm$0.24	&	0.21$\pm$0.07	&	4.59$\pm$0.01	&0.85$\pm$0.03	&	1.90	&	6.35E-27	&	78$\pm$2	&	3.03E-27	&	341$\pm$41	&&	7.61E-27	&	93$\pm$2	&	3.45E-27	&	388$\pm$33	\\
143018$^{5}$ & 0.56$^{+0.05}_{-0.05}$ &0.005&	3.10$^{4}$    	&	0.34		&	0.98	                	&	0.23	        	&	1.04		&	0.13		&	4.60	                	&	0.99		&	1.77	&	5.23E-27	&	64$\pm$2	&	2.73E-27	&	307$\pm$37	&&	6.49E-27	&80$\pm$2	&	3.15E-27	&	354$\pm$29	\\
143275$^{5}$ &	1.45$^{+0.28}_{-0.28}$	&0.49&	3.43$\pm$0.04$^{4}$	&	0.58$\pm$0.01	&	0.62$\pm$0.06	&0.25$\pm$0.01	&0.76$\pm$0.10	&	0.07$\pm$0.01	&	4.56$\pm$0.01	&	0.77$\pm$0.06	&	1.10	&	1.99E-27	&	24$\pm$2	&	1.01E-27&	113$\pm$56	&&	3.25E-27	&	40$\pm$2	&	1.43E-27	&160$\pm$51	\\
144217$^{5}$	&	1.37$^{+0.12}_{-0.12}$	&0.55&	3.68$\pm$0.03$^{4}$	&	0.70$\pm$0.01	&	1.29$\pm$0.04	&	0.10$\pm$0.01	&0.74$\pm$0.05	&	0.09$\pm$0.01	&4.50$\pm$0.01	&	0.66$\pm$0.03	&	1.42	&	3.40E-27	&	42$\pm$2	&	1.81E-27	&	204$\pm$54	&&	4.66E-27	&	57$\pm$2	&	2.23E-27	&	251$\pm$49	\\
147165$^{1}$	&	2.51$^{+0.51}_{-0.42}$	&0.58&	3.86$\pm$0.52$^{1}$	&	1.47$\pm$0.23	&	1.56$\pm$0.30	&	0.04$\pm$0.01	&	0.63$\pm$0.13	&	0.02$\pm$0.01	&	4.61$\pm$0.01	&	0.89$\pm$0.03	&	1.64	&	4.37E-27	&	54$\pm$2	&	2.31E-27	&	260$\pm$58	&&	5.63E-27	&	69$\pm$2	&	2.73E-27	&	307$\pm$53	\\
147933$^{1}$	&	5.01$^{+0.80}_{-0.80}$	&1.13&	5.74$\pm$0.40$^{1}$	&	2.58$\pm$0.34	&	1.23$\pm$0.17	&	0.02$\pm$0.01	&	0.58$\pm$0.09	&	0.10$\pm$0.01	&	4.58$\pm$0.01	&	0.95$\pm$0.03	&	1.51	&	3.74E-27	&	46$\pm$2	&	1.93E-27	&	217$\pm$58	&&	5.00E-27	&	61$\pm$2	&	2.35E-27	&	264$\pm$53	\\
149038$^{1}$	&	1.32	&-0.41&	3.12$\pm$0.15$^{5}$	&	1.00$\pm$0.05	&	1.29$\pm$0.26	&	0.29$\pm$0.07	&	1.99$\pm$0.53	&	0.22$\pm$0.08	&	4.58$\pm$0.01	&	0.99$\pm$0.03	&	2.72	&	1.02E-26	&	125$\pm$2	&	4.96E-27	&	558$\pm$66	&&	1.14E-26	&	140$\pm$2	&	5.38E-27	&	605$\pm$61	\\
149404$^{1}$	&	3.72	&0.17&	3.53$\pm$0.38$^{1}$	&	2.19$\pm$0.32	&	1.45$\pm$0.31	&	0.12$\pm$0.02	&	0.80$\pm$0.14	&	0.19$\pm$0.04	&	4.60$\pm$0.01	&	0.86$\pm$0.03	&	1.85	&	5.46E-27	&	67$\pm$2	&	2.80E-27	&	315$\pm$34	&&	6.72E-27	&	82$\pm$2	&	3.22E-27	&	362$\pm$24	\\
149757$^{1}$	&	1.40$^{+0.03}_{-0.03}$	&0.61&	3.09$^{8}$    	&	0.99$\pm$0.12	&	1.00$\pm$0.26	&	0.24$\pm$0.03	&	1.55$\pm$0.26	&	0.18$\pm$0.04	&	4.55$\pm$0.01	&	1.19$\pm$0.04	&	2.20	&	6.86E-27	&	84$\pm$2	&	3.69E-27	&	415$\pm$31	&&	8.12E-27	&100$\pm$2	&	4.10E-27	&	461$\pm$21	\\
151804$^{1}$	&	1.68	&-0.54&	3.77$\pm$0.16$^{5}$	&	1.13$\pm$0.12	&	1.45$\pm$0.38	&	0.13$\pm$0.03	&	0.62$\pm$0.13	&0.14$\pm$0.05	&	4.60$\pm$0.02	&0.76$\pm$0.02	&	2.13	&	7.95E-27	&	97$\pm$2	&	3.56E-27	&	400$\pm$68	&&	9.21E-27	&	113$\pm$2	&	3.97E-27	&	447$\pm$64	\\
152233$^{1}$	&	2.34	&-0.48&	2.95$\pm$0.29$^{1}$	&	1.33$\pm$0.24	&	0.54$\pm$0.15	&	0.36$\pm$0.05	&	0.96$\pm$0.21	&	0.10$\pm$0.03	&	4.61$\pm$0.03	&	0.99$\pm$0.03	&	2.07	&	7.48E-27	&	92$\pm$4	&	3.47E-27	&	390$\pm$67	&&	8.75E-27	&107$\pm$4	&	3.89E-27	&	437$\pm$63	\\
152236$^{1}$	&	6.92$^{+1.99}_{-1.55}$	&0.09&	3.73$\pm$0.39$^{1}$	&	2.24$\pm$0.26	&	0.76$\pm$0.66	&	0.26$\pm$0.05	&	1.29$\pm$0.37	&	0.15$\pm$0.03	&4.61$\pm$0.02	&	1.10$\pm$0.09	&	2.16	&	6.89E-27	&	84$\pm$3	&	3.66E-27	&	412$\pm$36	&&	8.15E-27	&100$\pm$3	&	4.08E-27	&	459$\pm$27	\\
152408$^{1}$	&	2.34	&-0.40&	3.81$\pm$0.13$^{5}$	&	1.60$\pm$0.20	&	1.45$\pm$0.24	&	0.11$\pm$0.02	&	0.65$\pm$0.12	&	0.18$\pm$0.04	&	4.58$\pm$0.01	&	0.79$\pm$0.03	&	2.11	&	7.43E-27	&	91$\pm$3	&	3.58E-27	&	402$\pm$65	&&	8.70E-27	&	107$\pm$3	&	3.99E-27 &449$\pm$61	\\
154368$^{1}$	&	3.89$^{+0.47}_{-0.42}$	&0.12&	3.33$\pm$0.15$^{1}$	&	2.53$\pm$0.20	&	1.09$\pm$0.10	&	0.22$\pm$0.01	&	1.05$\pm$0.09	&	0.22$\pm$0.02	&	4.58$\pm$0.00	&	1.00$\pm$0.02	&	2.00	&	6.16E-27	&	75$\pm$2	&	3.19E-27	&	359$\pm$35	&&	7.42E-27	&	91$\pm$2	&	3.61E-27	&	405$\pm$26	\\
164794$^{3}$	&	1.95	&-0.40&	3.57$\pm$0.17$^{5}$	&	1.18$\pm$0.06	&	1.31	 	&	0.10		&	1.11		&	0.13		&	4.56		&	0.99		&	1.73	&	5.74E-27	&	70$\pm$2	&	2.58E-27	&	290$\pm$53	&&	7.00E-27	&	86$\pm$2	&	3.00E-27	&	337$\pm$48	\\
193322$^{1}$	&	1.45	&-0.007&	2.78$\pm$0.26$^{1}$	&	1.11$\pm$0.15	&	0.91$\pm$0.28	&	0.30$\pm$0.04	&	1.09$\pm$0.17	&	0.03$\pm$0.01	&	4.58$\pm$0.01	&	0.92$\pm$0.03	&	2.13	&	6.84E-27	&	84$\pm$2	&	3.61E-27	&	406$\pm$41	&&	8.11E-27	&	99$\pm$2	&	4.03E-27	&	453$\pm$33	\\
199579$^{1}$	&	1.78$^{+0.31}_{-0.26}$	&-0.32&	3.17$\pm$0.69$^{1}$	&	1.14$\pm$0.28	&	1.10$\pm$0.55	&	0.28$\pm$0.06	&	0.81$\pm$0.21	&	0.20$\pm$0.06	&	4.59$\pm$0.01	&	0.99$\pm$0.03	&	2.07	&	7.06E-27	&	87$\pm$2	&	3.38E-27	&	380$\pm$63	&&	8.32E-27	&102$\pm$2	&	3.80E-27	&	427$\pm$58	\\
203064$^{1}$	&	1.39	&-0.25&	3.04$\pm$0.36$^{1}$	&	0.82$\pm$0.16	&	1.00$\pm$0.21	&	0.25$\pm$0.05	&	0.85$\pm$0.24	&	0.04$\pm$0.02	&	4.54$\pm$0.02	&	0.82$\pm$0.03	&	1.67	&	5.13E-27	&	63$\pm$2	&	2.37E-27	&	266$\pm$59	&&	6.40E-27	&	78$\pm$2	&	2.79E-27	&	313$\pm$54	\\
206267$^{1}$	&	2.88$^{+0.28}_{-0.25}$	&0.06&	2.82$\pm$0.16$^{1}$	&	1.47$\pm$0.14	&	1.17$\pm$0.36	&	0.27$\pm$0.02	&	1.02$\pm$	0.13	&0.22$\pm$0.04	&4.59$\pm$0.01	&	0.91$\pm$0.03	&	1.64	&	4.60E-27	&	56$\pm$2	&	2.42E-27	&	273$\pm$37	&&	5.86E-27	&	72$\pm$2	&	2.84E-27	&	319$\pm$29	\\
209339$^{1}$	&	1.82$^{+0.27}_{-0.23}$	&-0.27&	2.78$\pm$0.34$^{1}$	&	1.00$\pm$0.23	&	1.16$\pm$0.30	&	0.24$\pm$0.04	&	0.99$\pm$0.19	&	0.08$\pm$0.02	&	4.60$\pm$0.01	&	0.88$\pm$0.03	&	1.55	&	4.65E-27	&	57$\pm$2	&	2.24E-27	&	251$\pm$60	&&	5.91E-27	&	72$\pm$2	&	2.65E-27	&	298$\pm$55	\\
209975$^{1}$	&	1.80	&-0.15&	2.82$\pm$0.38$^{1}$	&	0.96$\pm$0.17	&	1.17$\pm$0.41	&	0.26$\pm$0.04	&	1.12$\pm$0.22	&	0.18$\pm$0.04	&	4.59$\pm$0.01	&	0.94$\pm$0.03	&	1.70	&	5.12E-27	&	63$\pm$2	&	2.50E-27	&	281$\pm$52	&&	6.38E-27	&	78$\pm$2	&	2.92E-27	&	328$\pm$46	\\
210839$^{1}$	&	2.95$^{+0.28}_{-0.26}$	&-0.06&	2.80$^{9}$	 	&	1.60$\pm$0.11	&	0.76$\pm$0.56	&	0.33$\pm$0.06	&	1.12$\pm$0.27	&0.13$\pm$0.03	&	4.60$\pm$0.02	&	0.96$\pm$0.06	&	1.54	&	4.28E-27	&	52$\pm$2	&	2.14E-27	&	240$\pm$44	&&	5.54E-27	&	68$\pm$2	&	2.55E-27	&	287$\pm$38	\\
303308$^{1}$ & 3.16 &-0.57& 3.02$\pm$0.21$^{1}$ & 1.36$\pm$0.18 & 0.87$\pm$0.21 & 0.26$\pm$0.03 &	0.90$\pm$0.13 & 0.16$\pm$0.03	& 4.59$\pm$0.01 & 0.95$\pm$0.03 & 1.23 & 4.03E-27 & 49$\pm$8	& 1.56E-27 & 175$\pm$68 && 5.30E-27 & 175$\pm$7 & 1.97E-27 &	222$\pm$64 \\
027778$^{2}$ & 2.29$\pm$1.20 &0.53& 2.79$\pm$0.38$^{2}$ & 1.09$\pm$0.03 & 0.87$\pm$0.71 & 0.32$\pm$0.01 & 1.31$\pm$0.10 & 0.24$\pm$0.01 & 4.59$\pm$0.04 & 1.21$\pm$0.03 & 1.39 & 3.08E-27 & 38$\pm$4 & 1.82E-27 & 205$\pm$42 && 4.34E-27 & 53$\pm$5 & 2.24E-27 & 252$\pm$35 \\
037061$^{2}$ & 5.37$\pm$1.23 &0.66& 4.29$\pm$0.21$^{2}$ & 2.40$\pm$0.21 & 1.54$\pm$0.00 & 0.00$\pm$0.00 & 0.31$\pm$0.00 & 0.05$\pm$0.00 & 4.57$\pm$0.00 & 0.90$\pm$0.00 & 1.19 & 2.28E-27 & 28$\pm$2 & 1.28E-27 & 144$\pm$46 && 3.55E-27 & 43$\pm$2 & 1.69E-27 & 190$\pm$39 \\
116852$^{1}$ & 1.05$\pm$1.20 &-1.15& 2.42$\pm$0.37$^{1}$ & 0.51$\pm$0.12 & 0.52$\pm$0.25 & 0.38$\pm$0.10 & 0.63$\pm$0.17 & 0.01$\pm$0.01 & 4.55$\pm$0.04 & 0.78$\pm$0.07 & 1.13 & 2.29E-27 & 28$\pm$2 & 1.09E-27 & 123$\pm$106 && 3.55E-27 & 44$\pm$2 & 1.51E-27 & 170$\pm$103 \\
122879$^{2}$ & 2.19$\pm$1.26 &-0.47& 3.17$\pm$0.20$^{2}$ & 1.41$\pm$0.04 & 1.15$\pm$0.08 & 0.15$\pm$0.02 & 0.53$\pm$0.06 & 0.09$\pm$0.01 & 4.57$\pm$0.04 & 0.74$\pm$0.02 & 1.50 & 7.89E-27 & 97$\pm$2 & 2.12E-27 & 239$\pm$98 && 9.15E-27 & 112$\pm$2 & 2.54E-27 & 286$\pm$95 \\
147888$^{2}$ & 5.89$\pm$1.20 &1.11& 4.08$\pm$0.18$^{2}$ & 1.97$\pm$0.03 & 1.43$\pm$0.05 & 0.03$\pm$0.01 & 0.74$\pm$0.07 & 0.07$\pm$0.01 & 4.58$\pm$0.03 & 0.94$\pm$0.02 & 0.95 & 1.21E-27 & 15$\pm$2 & 0.66E-27 & 74$\pm$36 && 2.47E-27 & 30$\pm$2 & 1.07E-27 & 121$\pm$27 \\
185418$^{2}$ & 2.57$\pm$1.17 &0.08& 2.67$\pm$0.20$^{2}$ & 1.39$\pm$0.04 & 1.37$\pm$0.09 & 0.16$\pm$0.02 & 1.89$\pm$0.12 & 0.14$\pm$0.01 & 4.59$\pm$0.03 & 1.03$\pm$0.02 &1.55 & 5.06E-27 & 62$\pm$6 & 2.00E-27 & 225$\pm$44 && 6.32E-27 & 77$\pm$8 & 2.42E-27 & 272$\pm$37 \\
207198$^{2}$ & 4.79$\pm$1.23 &0.40& 2.68$\pm$0.11$^{2}$ & 1.54$\pm$0.03 & 0.73$\pm$0.05 & 0.35$\pm$0.01 & 1.16$\pm$0.07 & 0.20$\pm$0.01 & 4.62$\pm$0.03 & 1.04$\pm$0.02 & 0.93 & 0.65E-27 & 8.0$\pm$2 & 0.58E-27 & 65$\pm$35 && 1.91E-27 & 24$\pm$2 & 1.00E-27 & 112$\pm$26 \\
\enddata 
\\
$^{a}$ Taken from Gudennavar et al.\ (2012) along 
       with upper and lower uncertainty
       (in unit of $10^{21}\cm^{-2}\HH$)\\
$^{b}$ Taken from Haris et al.\ (2016)\\

$^{c}$ The wavelength-integrated extinction  
       $\Aint \equiv\int_{912\Angstrom}^{10^{3}\um}
       A_\lambda/\NH\,d\lambda$
       (in unit of $10^{-25}\magni\cm^3\HH^{-1}$)\\
$^{d}$  The interstellar Si, Mg and Fe abundances 
        are taken to be that of proto-Sun (Lodders 2003)
        for which the total silicate volume per H atom 
        is $\Vsil/\rmH \approx 3.17\times10^{-27}\cm^{3}\HH^{-1}$.\\
$^{e}$  The interstellar Si, Mg and Fe abundances 
        are taken to be that of unevolved early 
        B stars (Przybilla et al.\ 2008)
        for which $\Vsil/\rmH \approx 2.52\times10^{-27}\cm^{3}\HH^{-1}$.\\
$^{f}$ The total volume of silicon dust per H atom
       (in unit of $\cm^3\HH^{-1}$) required to account 
       for the observed extinction.\\
$^{g}$ The silicon depletion $\stohd$
       (in unit of ppm) required to account 
       for the observed extinction.\\

$^{h}$ The total volume of carbon dust per H atom
       (in unit of $\cm^3\HH^{-1}$) required to account 
       for the observed extinction.\\
$^{i}$ The carbon depletion $\cdust$
       (in unit of ppm) required to account 
       for the observed extinction.
       
(1) Valencic et al.\ (2004); (2) Gordon et al.\ (2009); 
(3) Jenniskens \& Greenberg\ (1993); (4) Lewis et al.\ (2005);    
(5) Wegner et al.\ (2003); (6) Sofia et al.\ (2004);    
(7) Gnacinski \& Sikorski (1999);  (8) Cardelli et al.\ (1989);
(9) Rachford et al.\ (2008).

\end{deluxetable}

\clearpage

\thispagestyle{empty}
\setlength{\voffset}{25mm}
\begin{deluxetable}{lcccccccccccr}
\tablecolumns{24}
\tabletypesize{\scriptsize}
\tablewidth{0truein}
\center
\tablecaption{Model Parameters for Fitting
              the UV/Optical/Near-IR Extinction
              with A Mixture of Silicate/Graphite Dust
              \label{tab:modpara}
              }
\tablehead{
\colhead{Star}&
\colhead{$A_V/\NH$}&
\colhead{$\alpha_{\rm S}$}&
\colhead{$a_{c,{\rm S}}$}&
\colhead{$\alpha_{\rm C}$}&
\colhead{$a_{c,{\rm C}}$}&
\colhead{$\chi^2/{\rm dof}$}&
\colhead{$\cgas$}&
\colhead{$\cdust$}&
\colhead{$\sigas^{2}$}&
\colhead{$\sidust$}
\\
\colhead{}&
\colhead{($10^{-22}\magni\cm^{2}\,\rmH^{-1}$)}&
\colhead{}&
\colhead{($\um$)}&
\colhead{}&
\colhead{($\um$)}&
\colhead{}&
\colhead{(ppm)}&
\colhead{(ppm)}&
\colhead{(ppm)}&
\colhead{(ppm)}
}
\startdata
HD 001383&4.53&-3.0&0.18&-3.81&0.55&0.20&173$\pm$46$^{1}$&243&5.36$\pm$0.49&44\\
HD 002905&5.40&-3.3&0.35&-3.81&0.80&0.53&141$\pm$39$^{1}$&252&2.68$\pm$1.10&61\\
HD 023180&5.74&-2.6&0.16&-3.41&0.15&0.13&119$\pm$21$^{1}$&256&0.56$\pm$0.03&62\\
HD 024912&5.05&-2.8&0.14&-3.61&0.20&0.20&103$\pm$15$^{1}$&222&1.61$\pm$0.03&54\\
HD 030614&7.07&-2.9&0.20&-2.80&0.05&0.27&214$\pm$48$^{1}$&206&15.10$\pm$1.95&75\\
HD 034078&5.81&-2.5&0.14&-3.61&0.50&0.14&108$\pm$15$^{1}$&261&15.75$\pm$1.35&63\\
HD 037021&5.18&-2.0&0.14&-3.21&0.50&0.28&192$\pm$15$^{4}$&128&4.24$\pm$0.95&46\\
HD 041117&4.27&-3.2&0.23&-3.81&0.90&0.45&119$\pm$22$^{1}$&225&1.14$\pm$0.08&41\\
HD 042087&3.85&-3.3&0.22&-3.41&0.25&0.24&134$\pm$34$^{1}$&249&1.69$\pm$0.40&36\\
HD 045314&7.23&-2.9&0.20&-3.01&0.10&0.15&190$\pm$47$^{1}$&217&1.58$\pm$0.22&79\\
HD 046056&4.42&-3.1&0.10&-3.61&0.30&0.30&154$\pm$43$^{1}$&247&1.99$\pm$0.40&30\\
HD 046202&3.24&-3.2&0.24&-3.61&0.55&0.57&125$\pm$27$^{1}$&202&21.30$\pm$8.55&29\\
HD 069106&4.62&-2.4&0.10&-3.01&0.15&0.17&187$\pm$47$^{1}$&200&2.33$\pm$0.54&29\\
HD 093205&5.54&-3.1&0.27&-3.61&0.35&0.39&219$\pm$48$^{1}$&193&6.29$\pm$2.19&70\\
HD 093843&4.69&-3.1&0.41&-3.21&0.15&0.47&225$\pm$48$^{1}$&177&17.72$\pm$4.82&64\\
HD 101190&4.62&-3.2&0.14&-3.81&0.85&0.57&188$\pm$47$^{1}$&248&6.29$\pm$1.01&36\\
HD 112244&6.42&-2.9&0.33&-3.41&0.10&0.36&128$\pm$30$^{1}$&236&3.46$\pm$1.20&86\\
HD 143018&6.07&-2.9&0.16&-3.61&0.25&0.18&122$\pm$24$^{1}$&239&4.04$\pm$0.07&58\\
HD 143275&4.00&-3.4&0.33&-3.81&0.38&0.51&106$\pm$49$^{1}$&171&2.83$\pm$0.32&31\\
HD 144217&5.11&-2.6&0.18&-3.61&0.20&0.35&105$\pm$47$^{1}$&217&3.04$\pm$0.07&40\\
HD 147165&5.86&-2.0&0.14&-3.61&0.25&0.18&104$\pm$51$^{1}$&179&1.96$\pm$0.17&65\\
HD 147933&5.15&-2.3&0.27&-3.61&0.65&0.99&101$\pm$51$^{1}$&176&0.72$\pm$0.04&64\\
HD 149038&7.58&-2.8&0.18&-3.61&0.05&0.87&167$\pm$60$^{1}$&297&1.12$\pm$0.20&108\\
HD 149404&5.89&-3.0&0.37&-3.41&0.10&0.39&114$\pm$19$^{1}$&209&1.23$\pm$0.35&76\\
HD 149757&7.07&-2.8&0.14&-3.21&0.10&0.27&104$\pm$15$^{1}$&283&1.50$\pm$0.03&69\\
HD 151804&6.73&-2.9&0.45&-4.01&0.45&1.55&192$\pm$62$^{1}$&178&2.47$\pm$0.06&130\\
HD 152233&5.68&-3.0&0.14&-3.61&0.40&0.12&180$\pm$61$^{1}$&242&3.89$\pm$3.52&59\\
HD 152236&3.24&-3.0&0.24&-3.81&0.45&0.24&117$\pm$22$^{1}$&241&3.15$\pm$2.25&87\\
HD 152408&6.84&-2.9&0.49&-3.61&0.10&1.41&166$\pm$59$^{1}$&174&3.98$\pm$2.18&126\\
HD 154368&6.50&-3.1&0.31&-3.61&0.20&0.30&116$\pm$21$^{1}$&235&1.49$\pm$0.16&85\\
HD 164794&6.05&-2.5&0.16&-3.61&0.25&0.12&165$\pm$45$^{1}$&235&2.94$\pm$0.68&57\\
HD 193322&7.66&-2.7&0.14&-3.81&0.20&0.23&123$\pm$30$^{1}$&249&3.69$\pm$0.17&91\\
HD 199579&6.40&-3.1&0.25&-3.41&0.20&0.29&153$\pm$56$^{1}$&246&1.94$\pm$0.24&89\\
HD 203064&5.90&-3.0&0.26&-4.01&0.55&0.34&144$\pm$52$^{1}$&207&1.53$\pm$0.49&75\\
HD 206267&5.10&-3.1&0.20&-3.61&0.25&0.18&119$\pm$24$^{1}$&240&4.38$\pm$0.52&58\\
HD 209339&5.49&-2.4&0.08&-3.61&0.25&0.38&147$\pm$53$^{1}$&257&5.18$\pm$0.79&47\\
HD 209975&5.33&-2.9&0.18&-3.81&0.65&0.41&134$\pm$44$^{1}$&262&2.84$\pm$0.93&64\\
HD 210839&5.42&-3.0&0.12&-3.61&0.30&0.17&127$\pm$35$^{1}$&270&3.56$\pm$0.24&49\\
HD 303308&4.30&-2.9&0.12&-3.41&0.20&0.18&198$\pm$62$^{1}$&194&7.75$\pm$3.39&35\\
HD 027778$^{3}$&4.76&-3.1&0.14&-3.0&0.10&0.83&93$\pm$32$^{4}$&207&5$\pm$0.50&48\\
HD 037061$^{3}$&4.47&-2.0&0.17&-2.4&0.05&1.05&99$\pm$36$^{4}$&81&4$\pm$0.47&57\\
HD 116852$^{3}$&4.86&-2.7&0.08&-3.4&0.15&1.54&116$\pm$102$^{4}$&177&3$\pm$0.38&43\\
HD 122879$^{3}$&6.44&-2.7&0.18&-3.4&0.10&6.19&365$\pm$94$^{4}$&184&5$\pm$0.22&64\\
HD 147888$^{3}$&3.34&-2.1&0.16&-3.5&0.20&2.28&98$\pm$23$^{4}$&101&3$\pm$0.60&35\\
HD 185418$^{3}$&5.05&-2.0&0.06&-3.7&0.30&4.14&173$\pm$34$^{4}$&310&0.1$\pm$0.01&36\\
HD 207198$^{3}$&3.22&-2.8&0.06&-2.6&0.05&3.49&69$\pm$21$^{4}$&168&3$\pm$1.34&21\\

\enddata 
\\
$^{1}$ $\cgas$ estimated from $\nH$ 
       (see eq.\,\ref{eq:cgas}).\\
$^{2}$ $\sigas$ taken from Haris et al.\ (2016).\\
$^{3}$ Model parameters taken from Mishra \& Li\ (2015).\\
$^{4}$ $\cgas$ taken from Parvathi et al.\ (2012).\\
\end{deluxetable}

\end{document}